\documentclass[a4paper,fleqn,usenatbib]{mnras}

\usepackage{newtxtext,newtxmath}
\usepackage[export]{adjustbox}

\usepackage[T1]{fontenc}
\usepackage{ae,aecompl}

\usepackage{graphicx}	
\usepackage{amsmath}	
\usepackage{amssymb}	
\usepackage[english]{babel}
\usepackage{subfig}
\usepackage{tikz}
\usepackage{float}
\usepackage{natbib}
\usepackage{apalike}
\usepackage{wasysym}
\usepackage{ragged2e}
\usepackage{hyperref}

\title[Top-heavy IGIMF and dust in starbursts]{The influence of a top-heavy integrated galactic IMF and dust on the chemical evolution of high-redshift starbursts}

\author[Palla et al.]{
M. Palla$^{1,2}$\thanks{E-mail: marco.ball94@gmail.com},
F. Calura$^{3}$,
F. Matteucci$^{1,4,5}$,
X. L. Fan$^{6}$,
F. Vincenzo$^{7}$ and
E. Lacchin$^{3,8}$
\\
$^{1}$ Dipartimento di Fisica, Sezione di Astronomia, Universit{\'a} degli Studi di Trieste, via G. B. Tiepolo 11, I-34131, Trieste, Italy\\
$^{2}$ IFPU - Institute for Fundamental Physics of the Universe, Via Beirut 2, I-34014, Trieste, Italy\\
$^{3}$ INAF - OAS, Osservatorio di Astrofisica e Scienza dello Spazio di Bologna, 
via Gobetti 93/3, I-40129 Bologna, Italy\\
$^{4}$ INAF, Osservatorio Astronomico di Trieste, via G. B. Tiepolo 11, I-34131, Trieste, Italy\\
$^{5}$ INFN, Sezione di Trieste, via A. Valerio 2, I-34100, Trieste, Italy\\
$^{6}$ School of Physics and Technology, Wuhan University, 430072, Wuhan, Hubei, 430205, China\\
$^{7}$ Center for Cosmology and AstroParticle Physics, The Ohio State University, 191 West Woodruff Avenue, Columbus, OH 43210, USA\\
$^{8}$ Dipartimento di Fisica e Astronomia, Universit\'a di Bologna, via Gobetti 93/2, I-40129 Bologna, Italy\\
}

\date{Accepted 2020 March 24. Received 2020 March 24; in original form 2020 March 4.}

\pubyear{2020}

\begin{document}
\label{firstpage}
\pagerange{\pageref{firstpage}--\pageref{lastpage}}
\maketitle

\begin{abstract}
  We study the effects of the integrated galactic initial mass function (IGIMF) and dust evolution on the abundance patterns of high redshift starburst galaxies. In our chemical models, the rapid collapse of gas clouds triggers an intense and rapid star formation episode, which lasts until the onset of a galactic wind, powered by the thermal energy injected by stellar winds and supernova explosions. 
Our models follow the evolution of several chemical elements (C, N, $\alpha$-elements and Fe) both in the gas and dust phases. 
We test different values of $\beta$, the slope of the embedded cluster mass function for the IGIMF, where lower $\beta$ values imply a more top-heavy initial mass function (IMF). 
The computed abundances are compared to high-quality abundance measurements obtained in lensed galaxies and from composite spectra in large samples of star-forming galaxies in the redshift range $2 \lesssim z \lesssim 3$. 
The adoption of the IGIMF causes a sensible increase of the rate of star formation with respect to a standard Salpeter IMF, with a strong impact on chemical evolution.
We find that in order to reproduce the observed abundance patterns in these galaxies, either we need a very top-heavy IGIMF ($\beta < 2$) or large amounts of dust. 
In particular, if dust is important, the IGIMF should have $\beta \ge 2$, which means an IMF slightly more top-heavy than the Salpeter one. The evolution of the dust mass with time for galaxies of different mass and IMF is also computed, highlighting that the dust amount increases with a top-heavier IGIMF.
   
\end{abstract}





\defcitealias{Weidner11}{W11}
\section{Introduction}
\label{s:intro}

The stellar initial mass function (IMF) influences most observable
properties of stellar populations, as it 
regulates the relative fractions of low- and high-mass stars within them.\\
Massive stars (i.e. the stars with mass $>~8~M_{\odot}$) are known to be the main 
producers of $\alpha$-elements\footnote{Elements characterised by
capture of $\alpha$ particles. Examples are O, Mg, Si, S, Ca.} over
short ($\le30\, Myr$) timescales. On the other hand, the bulk of Fe
in a galaxy is known to be produced by Type Ia supernovae (SNe) over timescales
that can even reach or exceed the Hubble time 
\citep{MatteucciGreggio86,Matteucci01}.
Owing to these differences, chemical abundance ratios have been used as
powerful instruments for reconstructing the star formation history of galaxies. 

Besides 
chemical evolution, many other properties of a galaxy are strictly
related to the IMF, such as the present time stellar mass (\citealt{Kennicutt98}),
the integrated light of galaxies (\citealt{Conroy12}) as well as energetic feedback
from massive stars.  \\
At present, a complete theory able to explain the origin of the IMF does
not exist. Another fundamental issue yet to be clarified concerns the
universality of the IMF, as in principle in the local Universe it
could be different from high redshift galaxies
(e.g. \citealt{Larson98}), which are likely to be characterised by 
different physical conditions.\\

In this framework, a significant role is played by the integrated
galactic initial mass function (IGIMF) theory (e. g., \citealt{Kroupa03}; \citealt{Weidner05}).
It is based on a few
basic empirical evidences related to the birth of stars in local
star-forming environments, which include the fact that: (i) stars form in a clustered mode (\citealt{Lada03,Megheat16}), i.e. in groups of at least a few stars in the dense molecular cloud cores;  (ii) within each stellar cluster,
the IMF is observed to be universal and well approximated by a multiple power-law form (\citealt{Massey98}; \citealt{Pflamm07});
(iii) stellar clusters are distributed according to a single-slope power law (\citealt{Lada03}) and (iv) the upper mass end of the embedded cluster mass function  has been found to depend on the star formation rate (SFR) of the galaxy (\citealt{Weidner04}). The main consequence of these evidences is that the integrated IMF in disc galaxies (such as the Milky Way) is generally steeper than the stellar IMF within each single star cluster
(\citealt{Kroupa03}). \\
\citet{Weidner11} extended the IGIMF
theory to systems characterised by high star formation rates (SFR$>10
M_{\odot}yr^{-1}$), showing that in the most intensely star-forming
objects, i.e. in very massive and compact systems, the resulting IMF
becomes top-heavy, with extreme consequences on 
chemical enrichment and on stellar feedback. 

Various other studies indicated that a high-redshift
top-heavy IMF seems to be required to explain several properties of massive galaxies.
These properties include the observed evolution of the optical luminosity density (\citealt{Larson98}), the integrated [$\alpha$/Fe] ratios (\citealt{Calura09}; \citealt{DeMasi18}) and the colour-luminosity relation (\citealt{Gibson97}) in local spheroids, the observed galaxy
number counts in the infrared band and at submillimetric wavelengths (\citealt{Baugh05}), the isotopic ratios in high-z starbursts (\citealt{Romano17}; \citealt{Zhang18}) and the discrepancy between the observed present-day stellar mass density and the integral of the comoving SFR density \citep{Dave08}. In order to conciliate other indications in early type galaxies (e.g. \citealt{Cenarro03}; \citealt{Conroy12_2},\citeyear{Conroy12}; \citealt{LaBarbera13}) suggesting a bottom-heavy IMF, \citet{Weidner13} and \citet{Ferreras15} proposed also a time-dependent form of the IMF, switching from top-heavy during the initial burst of star formation to
bottom-heavy at later times.

Several evidences support the idea that local spheroids must
have experienced a starburst phase at high redshift.
This is inferred from the record of the their stellar populations,
in particular from their integrated ages and integrated abundances (e. g., \citealt{Matteucci94}; \citealt{spolaor10} and references therein). 
In principle, high-redshift starbursts might also 
present the physical conditions required for a top-heavy IMF, whose
signature might be encoded in their interstellar 
abundance pattern. For this reason, the turbulent, strongly pressurized ISM of starbursts represents an ideal
laboratory to probe the IMF in the progenitors of local spheroids, and in particular to test the hypothesis of a top-heavy IMF during the starburst phase. 

The investigation of the physical conditions of high-redshift starbursts as traced by their observed abundance pattern is the main motivation of the present paper. By means of chemical evolution models for proto-spheroids,
for the first time we aim at testing the effects of a top-heavy IGIMF on the chemical evolution of starbursts. \\ 
The results of our models will be compared with high-quality data from lensed high-redshift starbursts.  
Our models allow us to follow the evolution of the chemical abundances of
several species (C, N, $\alpha$-elements and Fe), both in the gas and in the dust phases.
In fact, our models can account for differential dust depletion that allows us to study the abundances of refractory elements (e.g. Mg, Fe) in such objects.
It is worth noting that the present paper is the first in which  a detailed treatment of dust is included in chemical evolution models with the IGIMF. 
The inclusion of dust will be particularly insightful, as 
previous chemical evolution models which did not take into account this ingredient failed to reproduce the [$\alpha$/Fe] ratios observed in similar objects, such as Lyman Break Galaxies (LBGs, \citealt{Matteucci02}; \citealt{Pipino11}).  

The present work follows various studies carried on in the last few years, aimed at assessing  the effects of the IGIMF on galactic chemical evolution in various environments characterised by different star formation histories, i.e. the solar neighbourhood (\citealt{Calura10}), dwarf galaxies (\citealt{Vincenzo15}, \citealt{Lacchin20}) and local elliptical galaxies (\citealt{Recchi09}; \citealt{DeMasi18}). \\
A novel formulation of the IGIMF was recently proposed by \citet{Yan17} and \citet{Jerabkova17,Jerabkova18} and firstly tested by \citet{Yan19} for elliptical galaxies. The implementation of such new formulation in models for starburst galaxies might be the subject of a future work.\\[0.2cm]

The paper is organized as follows:  in Section \ref{s:models}, we describe the IGIMF theory and chemical and dust evolution models adopted in this work. Our observational data are described in Section \ref{ss:obs_data}. Our results are presented in Section \ref{s:results}. Finally, conclusions are drawn in Section \ref{s:conclusion}.

\section{Models}
\label{s:models}

In this Section, we will describe the IGIMF 
theory and the main features of the chemical evolution models adopted in this paper.
We will describe the basic physical ingredients of the models, which are aimed at describing the 
starburst phase of massive proto-spheroids. Finally, we will outline the main assumptions
regarding the main processes regulating dust evolution, which also
have a strong influence on interstellar abundances. 

\subsection{IGIMF}
\label{ss:IGIMF}

Following the works of \citet{Kroupa03} and \citet{Weidner05}, the IGIMF is defined by weighting the canonical IMF, $\phi(m)$ (described later in this Section),
with the mass distribution of the stellar clusters
(called embedded cluster mass function, ECMF), $\xi_{ecl}(M_{ecl})$. The IGIMF theory starts from the assumption that star formation takes place in molecular cloud cores, i.e. in embedded stellar clusters.
The IGIMF $\xi_{\rm IGIMF}(m, t)$ can be expressed as a function of stellar mass $m$ and time $t$ as: 
\begin{multline}
\label{e:IGIMF_def}
    \xi_{\rm IGIMF}(m, t) =\\
=\int_{M_{\rm ecl,min}}^{M_{\rm ecl,max}(\psi(t))} \phi(m \le m_{\rm max}(M_{\rm ecl}))\,\xi_{ecl}(M_{ecl})\,dM_{\rm ecl},
\end{multline}
where  $M_{ecl}$ is the cluster mass.

The IGIMF is normalized as
\begin{equation*}
    \int_{m_{min}}^{m_{max}}  m \, \xi_{\rm IGIMF}(m, t) \,dm = 1. 
\end{equation*}
As can be seen from Equation \eqref{e:IGIMF_def},
the IGIMF adopted in this work has a time dependence, which is due to the SFR $\psi(t)$ of the parent galaxy,
following the model of \citet{Weidner11} (hereafter \citetalias{Weidner11}).
In the following, we will list the assumptions, based on empirical evidence, on which the IGIMF theory is based.

\begin{figure*}
\centering
\includegraphics[width=.7\textwidth]{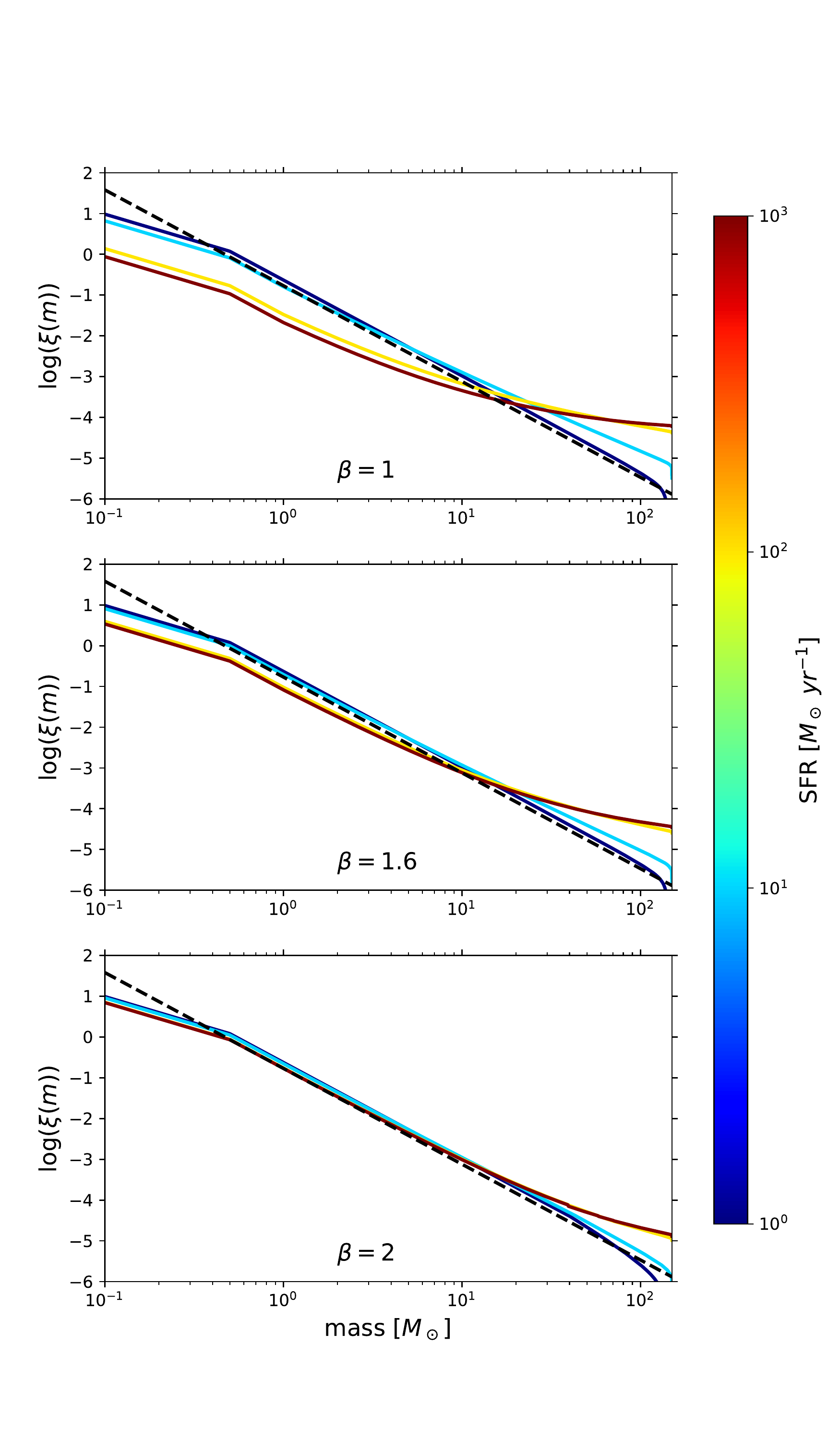}
\caption{Behaviour of the IGIMF adopted in this paper as a function of stellar mass and SFR for different
  values of $\beta$, namely the slope of the ECMF. 
  Upper panel: $\beta=1$; central panel: $\beta=1.6$; lower panel: $\beta=2$. In each panel, the four solid lines are the IGIMFs computed considering SFR=$1 M_\odot\,yr^{-1}$,$10 M_\odot\,yr^{-1}$, $100 M_\odot\,yr^{-1}$, $1000 M_\odot\,yr^{-1}$. The black dashed lines indicate the \citet{Salpeter55} IMF.} 
\label{f:IGIMF}
\end{figure*}

\begin{enumerate}

    \item The ECMF is represented by a single-slope power law:
    \begin{equation}
    \label{e:ECMF}
        \xi_{\rm ecl} (M_{\rm ecl}) \propto \bigg( \frac{M_{\rm ecl}}{M_{\rm ecl,max}} \bigg)^{-\beta},
    \end{equation}
    where the slope $\beta$ can vary between $\beta=0.5$ and $\beta=2.35$.\\
    The adopted minimum cluster mass is $M_{\rm ecl,min}=10^3M_\odot$ (\citetalias{Weidner11}).
This choice is due to
the fact that with high SFR values, the formation of low mass molecular cloud cores may
be suppressed due to the intense stellar feedback. 
However, for  $\beta\le 2$ (as adopted in this paper) the ECMF is not much sensitive to the adopted $M_{\rm ecl,min}$ value.

As for the upper mass limit $M_{\rm ecl,max}$, following \citet{Weidner04_} it can be expressed as: 
    \begin{equation}
    \label{e:maxmass_cluster}
M_{\rm ecl,max}=8.5 \cdot 10^4 \, \bigg(\frac{\psi(t)}{M_\odot\,yr^{-1}}\bigg)^{0.75} \, M_\odot,
    \end{equation}
    which holds for both low and high SFRs (\citealt{Bastian08}). We fix a maximum value for this upper mass limit at $10^7 M_\odot$, coherently with \citet{Weidner04_}.
    
    \item Within each embedded stellar cluster of a given mass $M_{\rm ecl}$, the IMF is assumed to be invariant. Following \citetalias{Weidner11}, we adopt the multi-component canonical IMF \citep{Kroupa01,Kroupa02}, which in its general form is expressed as:
    {\small
    \begin{equation}
    \label{eq:Kroupa_IMF}
\phi(m) = k \left\{\begin{array}{ll}
k^{'}\left(\frac{m}{m_{\rm H}} \right)^{-\alpha_{0}}&\hspace{-0.25cm} 0.01 \le m/M_\odot < 0.08 =m_{\rm H},\\
\left(\frac{m}{m_{\rm H}} \right)^{-\alpha_{1}}&\hspace{-0.25cm} 0.08 \le m/M_\odot < 0.50 =m_{0},\\
\left(\frac{m_{0}}{m_{\rm H}} \right)^{-\alpha_{1}}\left(\frac{m}{m_{0}} \right)^{-\alpha_{2}}&\hspace{-0.25cm}  0.50 \le m/M_\odot < 1.00 =m_{1},\\
\left(\frac{m_{0}}{m_{\rm H}} \right)^{-\alpha_{1}} \left(\frac{m_{1}}{m_{0}} \right)^{-\alpha_{2}}\left(\frac{m}{m_{1}} \right)^{-\alpha_{3}}&\hspace{-0.25cm} 1.00 \le m/M_\odot < m_{\rm max},\\
\end{array}\right.
\end{equation}
\normalsize\noindent with the following exponent values:
\small
\begin{equation*}
\alpha_0 = +0.30, \hspace{0.25cm}
\alpha_1 = +1.30, \hspace{0.25cm} 
\alpha_2 = +2.35, \hspace{0.25cm} 
\alpha_3 = +2.35.\\ 
\label{eq:imf}
    \end{equation*}}
$k$ and $k'$ are normalisation constants whereby the brown dwarf regime needs not be a continuous extension of the stellar regime (\citealt{Thies15}).
    In previous chemical evolution studies involving the IGIMF formalism, the quantity $\alpha_3$ has been kept equal to 2.35,
    independently from the cluster mass (e.g. \citealt{Recchi09,Calura10}).
As in  \citetalias{Weidner11}, for clusters with masses $M_{\rm ecl} > 2 \cdot 10^5 M_\odot$, the exponent $\alpha_3$ is parametrised as:
    \begin{equation}
        \alpha_3 (M_{\rm ecl}) = \left\{\begin{array}{ll}
        -1.67 \, \log_{10} \big( \frac{M_{\rm ecl}}{10^6 M_\odot} \big) + 1.05&\hspace{-0.25cm}  (M_{\rm ecl}\le 10^6 M_\odot),\\
       +1&\hspace{-0.25cm}  (M_{\rm ecl} > 10^6 M_\odot).\\
        \end{array} \right.
    \end{equation}

 The upper stellar mass limit $m_{\rm max}$ is computed from the mass of the embedded cluster $M_{\rm ecl}$,
 but in any case is always assumed $\le 150 M_\odot$ (see \citealt{Weidner04} for more details).
\end{enumerate}

In the present study, a metallicity dependence of the IGIMF is not taken into account (see  \citealt{Recchi14}; \citealt{Vincenzo15}; \citealt{Yan19}).
The adoption of a metallicity dependent IGIMF  (\citealt{Jerabkova18}) in high redshift starburst galaxies might be the subject of a future work. 

    \subsubsection{IGIMF behaviour as a function of $\beta$ and SFR}
    
In Figure \ref{f:IGIMF}, we show the IGIMF obtained with our prescriptions and for different values of the SFR, in which 
we have selected three values of $\beta$ among those adopted in \citetalias{Weidner11}: $\beta=1$, $\beta=1.6$ and $\beta=2$.

We do not consider the most extreme values such as $\beta=0.5$ and $\beta=2.35$. By adopting an ECMF with $\beta=1$ we obtain, for $\psi\gtrsim10 M_\odot yr^{-1}$,
an IMF comparable to the single-slope IMF of \citet{Gibson97}, a quite extreme top-heavy one (characterised by an index $x=0.8$ ($x=\alpha-1$, where $\alpha=2.35$ is the \citet{Salpeter55} IMF index) over the whole stellar mass range).
On the other hand, the IGIMF obtained adopting $\beta=2$ is very similar to the \citet{Salpeter55} IMF over most of the stellar range,
except at very high SFR values ($> 100 M_\odot yr^{-1}$). 
    
The IGIMFs calculated at low SFR values ($1 M_\odot\, yr^{-1}$) show a uniform decline with mass and the shape of a double-power law, with a knee located at $0.5~M_{\odot}$. 
A cut-off is visible at mass values larger than $\sim 100 M_{\odot}$, where the decrease is steeper and where the behaviour is similar to the IGIMF shown in, e. g., \citet{Recchi09}. 

In general, the higher the SFR value, the flatter IGIMF, the higher the relative number of massive stars, as due to increasing $M_{ecl,max}$ values with increasing SFR. 
Moreover, the lower the $\beta$ value, the stronger the IGIMF dependence on the SFR.

\subsection{Chemical evolution model}
\label{ss:chem_evo}
The model used in this work was originally designed to study the evolution of elliptical galaxies (\citealt{Matteucci94}; \citealt{Pipino11}; \citealt{Calura14}; \citealt{DeMasi18}).
The model has been improved by including the formation, growth and destruction of dust grains, following \citet{Gioannini17}.
This allows us to follow the evolution of the abundances of refractory elements (e.g. Si, Fe) in the gas and dust. 


\subsubsection{Chemical evolution model}
In our scheme, elliptical galaxies form from the rapid collapse of a gas cloud with primordial chemical composition, described by an exponential infall law. The galaxy is allowed to evolve as an $\lq$open box' into the potential well of a dark matter halo. The initial rapid collapse triggers an intense and rapid star formation (SF) episode, i.e. a starburst, which lasts until a galactic wind, powered by the thermal energy injected by stellar winds and SN explosions, occurs.
After that time, the galaxy evolves passively, i.e. with no more SF. This is a good approximation as real early-type galaxies have less than 1 per cent of their stellar population younger than 3 Gyr (\citealt{Salvador19}).\\

In this scenario, the evolution of a given chemical element $i$ is described by:
\begin{equation}
    \label{e:chem_evo}
    \dot{G_{i}}=-\psi(t)X_{i}(t) + R_{i}(t) + (\dot{G_{i}})_{inf} -
    (\dot{G_{i}})_{out}
    \end{equation}
where $G_{i}(t)=X_{i}(t)\,G(t)$ is the gas mass in the form of an element $i$ normalised to the total baryonic mass $M_{lum}$ and $G(t)=M_{gas}(t)/M_{lum}$ is the fractional mass of gas present in the galaxy at the time $t$. The quantity $X_{i}(t)$  represents the abundance fraction in mass of a given element $i$, with the summation over all elements in the gas mixture being equal to unity.

$R_{i}(t)$ represents the returned fraction of matter in the
form of an element $i$ that the stars eject into the interstellar medium (ISM). This term contains all the nucleosynthesis prescriptions about single low-intermediate mass stars (LIMS, $m<8 M_{\odot}$), core collapse (CC) SNe (Type II and Ib/c, $m>8 M_{\odot}$) and Type Ia SNe, for which we assume the single-degenerate (SD) scenario. In this scenario, a C-O white dwarf in a binary system accretes mass from a non-degenerate companion until it reaches the Chandrasekhar mass ($\sim1.44 M_{\odot}$) and explodes via C-deflagration.
The stellar yields are taken from \citet{VDH97} (LIMS), \citet{Francois04} (revised version of \citealt{WW95}, for massive stars; for nitrogen see \ref{sss:volatiles}) and \citet{Iwamoto99} (Type Ia SNe). The fraction of stars in binary systems able to originate Type Ia SNe is fixed at a value able to reproduce the present day Ia SN rate observed in local ellipticals (\citealt{Matteucci01}; \citealt{Calura06}; \citealt{Pipino11b}).\\
In Equation \ref{e:chem_evo}, the term $(\dot{G_{i}})_{inf}$ accounts for the infall of external gas. As for the infall, we assume an exponential law:
\begin{equation}
    (\dot{G_{i}})_{inf}\propto X_{i,inf} \exp{(-t/\tau_{inf})}, 
\end{equation}
where $X_{i,inf}$ describes the chemical composition of the infalling gas, assumed to be primordial.  The quantity $\tau_{inf}$ is the infall timescale.\\ 
The last term of Equation \eqref{e:chem_evo} represents the galactic wind. The occurrence of the wind is determined
by the condition that the thermal energy of the ISM, as due to feedback from SNe and stellar winds, is larger than or equal to the gas binding energy. The feedback prescriptions assumed here are the same as in \citet{DeMasi18}: in particular, we assume that only a small, variable fraction (generally of the order of a few percent) of the initial blast wave energy of CC-SNe, $E_o= 10^{51}$erg, is deposited in the ISM (see \citealt{Pipino02,Pipino04} for details), whereas all the initial blast wave energy of Type Ia SNe (the same as for CC-SNe) is restored into the ISM, as suggested by \citet{Recchi01}: in fact, when Type Ia SNe explode, the ISM is already hot beacause of the explosion of CC-SNe. 
Moreover, we assume that stellar winds by massive stars can inject into the ISM 3\% of the typical energy of stellar winds ($\sim 10^{49}$ erg, see \citealt{Bradamante98} for details). \\
The dark matter halo is assumed ten times more massive than the luminous mass, with its core radius being ten times larger than the effective radius (see \citealt{Matteucci94}; \citealt{Pipino04}; \citealt{DeMasi18}). The large core radius adopted is suggested by observed galaxies but cannot be generated in self consistent dark matter based models of galaxy formation. The value used here can also be interpreted as the natural core radius of Milgromian potentials, i.e. the phantom dark matter potential generated by the baryonic component (e.g. \citealt{Lueghausen13,Lueghausen15}).
The evolution of the binding energy $E_{bind}$ and thermal energy $E_{th}$ of the gas
as a function of time for all our models is shown in the bottom-right panels of figures \ref{f:summary10}, \ref{f:summary11} and \ref{f:summary12}.
In each figure, the binding energy (solid line) is computed from the amount of gas and dark matter available in the galaxy, 
whereas the thermal energy (dashed line) is the cumulative energy deposited by stellar winds and SNe in the gas.\\
\begin{table}
\Centering
\caption{Main parameters assumed for our chemical evolution models for starburst galaxies.} 
\begin{tabular}{l|llll}
\noalign{\smallskip}
\hline
\hline
\noalign{\smallskip}
 Model name            &   $M_{lum}$      &  $R_{eff}$     & $\nu$          & $\tau_{inf}$       \\      
                       &  $(M_{\odot})$   &   (kpc)       &  (Gyr$^{-1}$)   &  (Gyr)             \\    
\noalign{\smallskip}                                                                                                
\hline                                                                                                              

M3E10                  & $3 \times 10^{10}$  & 2         &   5             &    0.5           \\              
M1E11                  & $1 \times 10^{11}$  & 3         &   10            &    0.4           \\              
M1E12                  & $1 \times 10^{12}$  & 10        &   20            &    0.2           \\              
\hline
\hline
\end{tabular}
\label{t:models}
\begin{flushleft}
\end{flushleft}
\end{table}
The SFR is calculated as:
\begin{equation}
    \psi(t) = \nu G(t), 
\end{equation} 
i.e. it is assumed to be proportional to the gas mass via a constant $\nu$, the star formation efficiency, according to the Schmidt-Kennicutt law (\citealt{Schmidt59}; \citealt{Kennicutt98}). As in the $\lq\lq$inverse wind model" of \citet{Matteucci94}, the star formation efficiency is allowed to vary, increasing with galactic mass. This allows to reproduce the $\lq\lq$downsizing" behaviour of galaxies, with a a galactic wind occurring at earlier times in more massive systems, thus producing higher $\alpha$-elements relative to Fe abundance ratios in the stellar populations of the most massive galaxies, in agreement with observations (see also \citealt{DeMasi18}).\\
In the remainder of the paper, we will compare the results obtained with the IGIMF described in Section \ref{ss:IGIMF}  with those obtained with a standard \citet{Salpeter55} IMF, expressed by a single power law as $\xi(m) \propto m^{-2.35}$. 

The main features of the models used in this paper are summarised in Table \ref{t:models}. In the first column, the name of the model is shown.  The second column shows the adopted total baryonic mass. The third, the fourth and the fifth columns indicate for each model the adopted effective radius, the star formation efficiency and the infall timescale, respectively.

\subsubsection{Dust evolution model}
\label{sss:dust_model}

The chemical evolution model also follows in detail the various processes (production, growth, destruction) that influence dust evolution. 
Here we adopt the same formalism as used in previous works on chemical evolution models with dust (e.g. \citealt{Dwek98}; \citealt{Calura08}; \citealt{Vladilo18}; \citealt{Palla20}).\\

For  a  specific  element $i$ in the dust phase we have:
\begin{equation}
    \begin{split}
\dot{G}_{i,dust}= - \psi(t) X_{i,dust}(t) + \delta_i R_{i}(t) + {G}_{i,dust}(t)/\tau_{i,accr} +\\
-{G}_{i,dust}(t)/\tau_{i,destr} - \dot{G}_{i,dust,w}(t) \hspace*{1.6cm},
\label{eq:dust}
\end{split}
\end{equation}
where $G_{i,dust}$ and $X_{i,dust}$ are the normalised dust mass and the abundance in dust in the form of an element $i$, respectively. 

Equation~\ref{eq:dust} includes dust  production  from  AGB  stars and CC-SNe (expressed by the rate $\delta_i R_i$), dust growth from refractory elements in the gas phase in the cold ISM (with a rate $G_{i,dust}/\tau_{i,accr}$) and
destruction  by  SNe forward shocks (expressed by $G_{i,dust}/\tau_{i,destr}$).

As for dust production, we use the metallicity-dependent prescriptions from \citet{Bianchi07} for CC-SNe and \citet{DellAgli17} for AGB stars.
We also test the role of the reverse shock in the dust yields of CC-SNe.
In SN ejecta, the role of the reverse shock in the evolution of the dust deserves
prticular attention, as in some studies it may destroy large amounts of the dust mass initially produced (e.g.,  \citealt{Bianchi07}). 
To better investigate this aspect, we run a few models which include the reverse shock, and a few ones which do not include it. \\
At variance with previous works on chemical evolution in elliptical galaxies \citep{Pipino11,Grieco14}, we assume that Type Ia SNe do not produce dust.
This assumption is supported both by theoretical and observational arguments \citep{Nozawa11,Gomez12}. \\
Concerning the processes of growth and destruction, we calculate the metallicity-dependent timescales $\tau_{i,accr}$, $\tau_{i,destr}$ as in \citet{Asano13}.
$\tau_{i,accr}$ depends on the temperature $T$, density $n_H$, metallicity $Z$ and grain size $a$:
\begin{equation}
\tau_{i,accr}\propto \frac{1}{a\cdot n_H \cdot Z \cdot T} ,
\end{equation}
where we assume the reference values of $n_H=100cm^{-3}$ (dust condensation takes place in dense gas), $a=0.1\mu m$, $T=50K$ (as suggested by \citealt{Asano13}).
As for destruction by SNe, instead, we have:
\begin{equation}
\tau_{i,destr}\propto \frac{1}{\epsilon \cdot M_{swept} \cdot SN_{rate}}, 
\end{equation}
where we assume an efficiency of destruction $\epsilon=0.1$ (as suggested by \citealt{Asano13}) and a swept mass by SN forward shock $M_{swept}$ dependent on metallicity $Z$.

Because of the uncertainties related to dust growth in the ISM, we also run models without any dust growth in the ISM. In fact, it was shown that dust condensation in high redshift galaxies can encounter theoretical problems \citep{Ferrara16}, as the large amount of UV radiation from massive stars in starbursts can impact on the net dust growth rate \citep{Gall18}. On the other hand, grain growth in the ISM is also required to explain the large depletion rates of Fe, whose main producers (Type Ia SNe) do not seem to contribute to dust production (e.g. \citealt{Nittler18}). \\
At the same time, we also test models without forward shock destruction by SNe. In fact, the impact of these phenomena on the dust survival rate is still a matter of debate (\citealt{Gall18} and references therein).\\
In our plots, the results of our models which include the various processes of dust production, growth and destruction are dubbed as follows. All the models which include reverse shocks in SNe are labelled with 'R', and models in which reverse shock is turned off are labelled 'NR'.
Models which include growth (accretion) and destruction are labelled 'A' and 'D', respectively.
As examples, a model which includes reverse shock, growth and destruction will be 'ADR', whereas a model in which growth and destruction are absent but reverse shock is present will be dubbed 'R'.

\begin{table*}
    \centering
    \caption{Main features of the sample of starburst galaxies included in our sample. Objects in the upper part of the table (i.e. above the horizontal line) are lensed galaxies.
      Objects below the line are stacked spectra of galaxies.}
    \begin{tabular}{l|llllll}
    \noalign{\smallskip}
    \hline
    \hline
    \noalign{\smallskip}
 Object       &   Redshift    &   SFR        &   $M_{*}$      &  Notes      &    References\\      
              &               &   ($M_{\odot}/yr$)   &   ($M_{\odot}$)& &             &   \\    
              
    \noalign{\smallskip}                                                          
    \hline                                                                     
MS 1512-cB58  &  2.7276    &   $\sim$25-$\sim$150       &   $\sim$10$^{10}$      &   (1), (2), (3)    &    \citeauthor{Pettini00}(\citeyear{Pettini00}, \citeyear{Pettini02}); \citet{Teplitz00}; \citet{Siana08}\\
 8 o'clock arc  &  2.7350    &   $\sim$270       &   $\sim$10$^{11.6}$      &  (2), (3)    &    \citet{Dessauges10};  \citet{Finkelstein09}\\
 Cosmic Horseshoe  &  2.3812    &   $\sim$95-$\sim$190     &   $\sim$10$^{10}$      &  (1), (2), (3)    &    \citet{Quider09}; \citet{Hainline09} \\
SGAS J105039.6+001730  &  3.6252    &   $\sim$90-$\sim$140 &   $\sim$10$^{9.7}$    &   (2)   &    \citet{Bayliss14}   \\
RCSGA 032727-132609  &  1.7037   &   $\sim$130-$\sim$360       &   $\sim$10$^{10.3}$   &   (2)    &    \citet{Wuyts10}; \citet{Rigby11}   \\
SMACS J0304.3-4402  &  1.96    &   $\sim$20-$\sim$90        &   $\sim$10$^{10.8}$   &  (2)    &    \citeauthor{Christensen12a}(\citeyear{Christensen12a}, \citeyear{Christensen12b})    \\
SMACS J2031.8-4036  &  3.51    &   $\sim$15-$\sim$30 &   $\sim$10$^{9.4}$      &   (2)  &    \citeauthor{Christensen12a}(\citeyear{Christensen12a}, \citeyear{Christensen12b})    \\
    \hline
KBSS-LM1 Composite   &  2.396$\pm$0.111   &   $\sim$50-$\sim$55  &   $\sim$10$^{10}$   &  (2), (4)     &  \citet{Steidel16}\\
Shapley LBG Composite   &  $\sim$3   &   $>$50  &   -   &  (2)       &        \citet{Shapley03}; \citet{Pettini01} \\
    \hline
    \hline
    \end{tabular}
    \\[0.1cm]
    SFR and $M_*$ estimates were derived adopting a \citet{Salpeter55} IMF.\\
    (1) $M_*$ is the baryonic mass; (2) abundances from emission lines; (3) abundances from absorption lines; (4) median values of the sample, except for redshift (medium value with rms of the sample).
    \label{t:data}
\end{table*}

\begin{table*}
    \centering
    \caption{Abundance ratios of starburst galaxies included in our sample. Objects above the line are lensed galaxies. Objects below the line are stacked spectra of galaxies. In parentheses, the method of estimation for $\log$(O/H)+12 is indicated.}
    \begin{tabular}{l|lllllll}
    \noalign{\smallskip}
    \hline
    \hline
    \noalign{\smallskip}
 Object       &   log(O/H)+12    &   log(N/O)        &  log(C/O)      &    [Fe/H]$^1$    &  [Si/H]$^1$      &	[Mg/H]$^1$    \\       
              
    \noalign{\smallskip}                                                          
    \hline                                                                     
MS 1512-cB58  &  8.39$\pm$0.10 ($R_{23}$)   &   -1.24$\pm$0.14 \citep{Teplitz00} &   -   &  -1.15$\pm$0.1  &  -0.37$\pm$0.1 &    -0.32$\pm$0.1 \\
 & & -1.89$\pm$0.14 \citep{Pettini02} & & & &   \\
8 o'clock arc  &  8.58$\pm$0.18 ($N_2$)  &  -   &  -  &  -0.93$\pm$0.15 &  -0.19$\pm$0.14  & -\\
Cosmic Horseshoe  &  8.38$\pm$0.18 ($N_2$)  &  -   & -  &  -1.17$^{+0.18}_{-0.15}$ &  -0.29$^{+0.18}_{-0.15}$  & -\\
SGAS J105039.6+001730  &  $\geq$8.05 (direct)   &  -1.59$\pm$0.2   & -0.79$\pm$0.06  & -  &  -   &  - \\
 & 8.17$\pm$0.12 ($R_{23}$) & & & & &   \\
RCSGA 032727-132609  &  $\geq$8.14 (direct)   &  1.7$\pm$0.02  & -  &  -  & -  &  -    \\
 & 8.20$\pm$0.08 ($R_{23}$) & & & & &   \\
 & 8.20$\pm$0.04 ($N_{2}$) & & & & &   \\
SMACS J0304.3-4402  & 8.07$\pm$0.09 (direct)   &  -1.64$\pm$0.05  &  -  & -  & -  & - \\
 & 8.16$\pm$0.01 ($R_{23}$) & & & & &   \\
SMACS J2031.8-4036  &  7.76$\pm$0.03 (direct)  &  -  &  -0.80$\pm$0.09   &  -  &  -  & -  \\
 & 7.74$\pm$0.03 ($R_{23}$) & & & & &   \\
    \hline
KBSS-LM1 Composite   &  8.38$\pm$0.05 (direct)   &  -1.24$\pm$0.04   &  -0.60$\pm$0.09   & -   &  -  &  -\\
 & 8.20$\pm$0.1 ($R_{23}$) & & & & &   \\
  & 8.32$\pm$0.05 ($N_{2}$) & & & & &   \\
Shapley LBG Composite   &  [7.64,8.73] ($R_{23}$)  & -  &  [-0.81,-0.56]  & -  & -  & -  \\
    \hline
    \hline
    \end{tabular}
    \\[0.1cm]
   $^1$ [X/Y]=$\log(X/Y)-\log(X_\odot/Y_\odot)$, where $X$, $Y$ are abundances in the ISM for the object studied and $X_\odot$, $Y_\odot$ are solar abundances (from \citealt{Asplund09}).
    \label{t:abundance_data}
\end{table*}



\section{Observational data}
\label{ss:obs_data} 
In Table \ref{t:data} and \ref{t:abundance_data} we list the observed systems selected for this study with their main features
(redshift, SFR, stellar mass) and their chemical abundances. 

In our analysis, we consider several lensed objects (LBGs and Ly-$\alpha$ emitters) (upper parts of Tables \ref{t:data} and \ref{t:abundance_data}). 
The magnification due to gravitational lensing allows one to perform high quality spectroscopy, from which physical properties and chemical abundances can be derived from high S/N data (\citealt{Bayliss14} and references therein). \\
We extend our sample of  high-redshift star-forming systems  by considering also two cases where $\lq$composite' spectra from sizable sets of high-z systems were obtained with observations in non-lensed fields.
The data for composite samples are reported in the lower part of Tables \ref{t:data} and \ref{t:abundance_data}. 
These samples include high-redshift systems, with average stellar masses and SFR values comparable to the ones of the lensed objects.  
In the following, we will give a brief description of each of the objects considered in our analysis.

\subsection{MS 1512 - cB58}
MS 1512-cB58 (hereafter, cB58) is a lensed LBG first discovered by \citet{Yee96}, with redshift $z=2.7276$ \citep{Pettini02}. It is magnified by a factor $\sim 30$ by the cluster MS 1512+36 at $z=0.37$ \citep{Pettini00}.\\ Several SFR estimates were performed for this object in the past years, leading to different results depending on the adopted SFR estimator (e.g. IR, H$\alpha$, UV).  The available values span the range $\sim 25- \sim 150 M_\odot yr^{-1}$ \citep{Siana08}.
As noted by \citet{Siana08}, however, the highest SFR values might represent upper limits due to overestimated extinction corretions. \\
The estimated baryonic mass is of $\sim 10^{10} M_\odot$ (\citealt{Pettini00,Baker04}), whereas the effective radius is $R_{eff}\sim 2 kpc$ \citep{Seitz98}.\\
The observed abundances in Table \ref{t:abundance_data} are from \citet{Teplitz00} and \citet{Pettini02}. 
The 12+ log(O/H) and log (N/O) abundances were estimated by  \citet{Teplitz00} from interstellar emission lines. The O abundance was calculated by means of the $R_{23}$ indicator. 
The abundances obtained by \citet{Pettini02} were computed from interstellar absorption lines, and in particukar from the measured column densities using the apparent optical depth method. 
The N/O values from both sets will be compared with the models, in order to appreciate the differences in the measures as obtained from emission and absorption lines.

\subsection{8 o'clock arc}
The 8 o'clock arc is a LBG at redshift $z=2.7350$, lensed by the luminous red galaxy (LRG) SDSS J002240.91+143110.4 at $z=0.38$ \citep{Allam07}. The arc has an inferred stellar mass of $4.2\cdot 10^{11}M_\odot$ and a SFR of $266\pm74 M_\odot yr^{-1}$ (corrected for a lensing magnification of $\mu=8$, \citealt{Finkelstein09}). Analysis of the Baldwin-Phillips-Terlevich (BPT) diagram excludes substantial AGN contamination \citep{Finkelstein09}.\\
In Table \ref{t:abundance_data}, we show the O abundances from the Keck/LRIS emission spectrum (\citealt{Finkelstein09}) and the Fe, Si abundances obtained from the VLT/X-Shooter absorption spectrum of \citet{Dessauges10}.
O abundances are obtained from the $N_2$ indicator, whereas Fe, Si abundances by means of the apparent optical depth method (as in \citealt{Pettini02}).
  
\subsection{Cosmic horseshoe}
The Cosmic Horseshoe is a gravitationally lensed LBG discovered by \citet{Belokurov07}, with redshift $z=2.3812$. It is magnified by a factor $\mu=24\pm2$ by a massive LRG at $z=0.444$ \citep{Dye08}.
The inferred baryonic mass and effective radius are of the order of $10^{10}M_\odot$ and $2.5 kpc$ \citep{Hainline09}, respectively.
Its $H_{\alpha}$ and UV luminosities yield SFR values of $95 M_\odot yr^{-1}$ \citep{Quider09} and $190 M_\odot yr^{-1}$ \citep{Hainline09}, respectively.\\ 
The absorption lines from its Keck II/ESI spectrum (\citealt{Quider09}) and the emission lines from Keck II/NIRSPEC (\citealt{Hainline09})
give the abundances presented in Table \ref{t:abundance_data}. The abundances for Fe and Si were obtained by means of the apparent optical depth method, whereas its O abundance from the $N_2$ indicator. 

\subsection{SGAS J105039.6+001730}
The SGAS J105039.6+001730 LBG at redshift $z=3.6252$ is a system lensed  by a foreground galaxy cluster at $z=0.593$.
The lensing magnification is of the order of $\sim 30$. By taking into account the magnification factor, \citet{Bayliss14}
derived for the galaxy a stellar mass of $5\cdot 10^9 M_\odot$ and values of the SFR between $90$ and $140M_\odot yr^{-1}$ (depending if based on [OII] or $H_\beta$ luminosity, respectively).
The Magellan/FIRE interstellar emission spectrum ruled out a substantial AGN contribution. \\
Abundances in Table \ref{t:abundance_data} are taken from Magellan/FIRE spectrum \citep{Bayliss14}. As for the O abundance, both the direct measure and the value inferred via the $R_{23}$ indicator are presented. 

\begin{figure*}
\centering
\includegraphics[width=.78\textwidth]{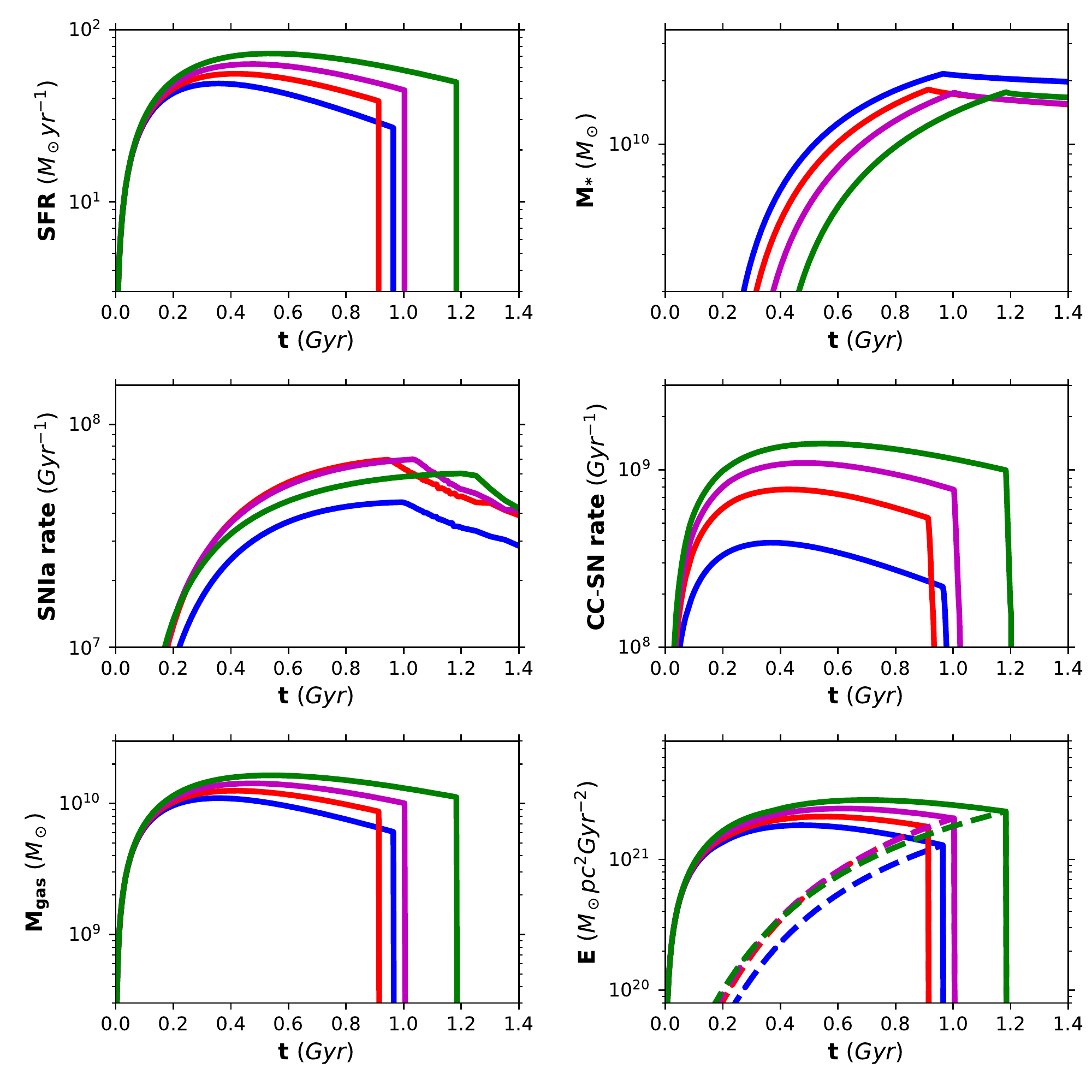}
\caption{From top-left corner, clockwise: time evolution of the SFRs, stellar mass, CC-SN rate, energetic budget, gas mass and Type Ia SN rates obtained for the M3E10 model (see Table \ref{t:models}) with a \citet{Salpeter55} IMF (blue lines) and \citetalias{Weidner11} IGIMF calculated for $\beta=1$ (green lines), $\beta=1.6$ (magenta lines) and $\beta=2$ (red lines). The sharp truncation in the SFR, CC-SN rate and gas mass are due to the onset of the galactic wind,
  which devoids the galaxy from the residual gas. In the enegetic budget plot, the solid lines represent the binding energy and the dashed lines the thermal energy in
all the different models. } 
\label{f:summary10}
    \end{figure*}

    \begin{figure*}
\centering
\includegraphics[width=.78\textwidth]{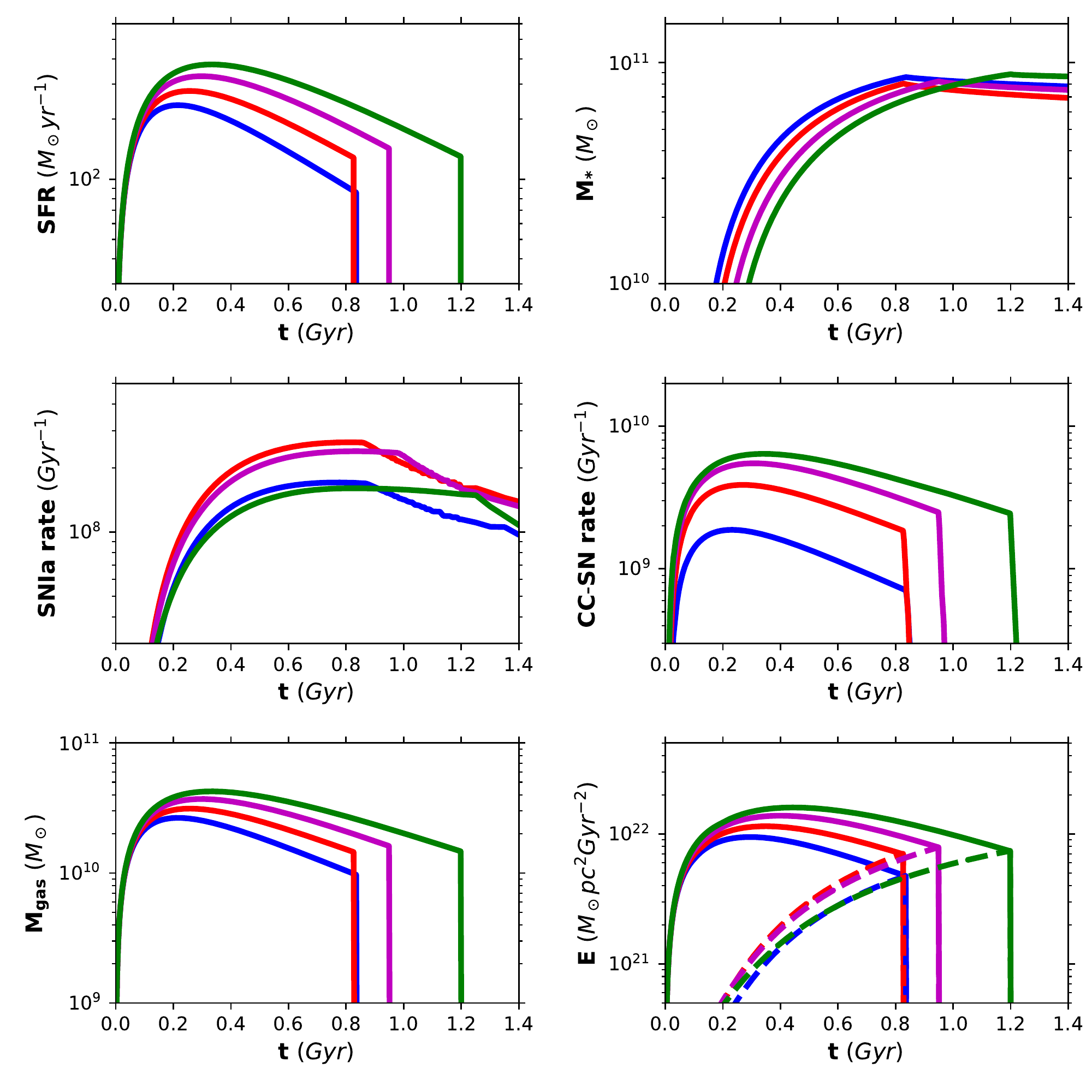}
\caption{Lines are as in Fig. \ref{f:summary10}, but computed for the M1E11 model of  Table \ref{t:models}.} 
\label{f:summary11}
    \end{figure*}

    \begin{figure*}
\centering
\includegraphics[width=.78\textwidth]{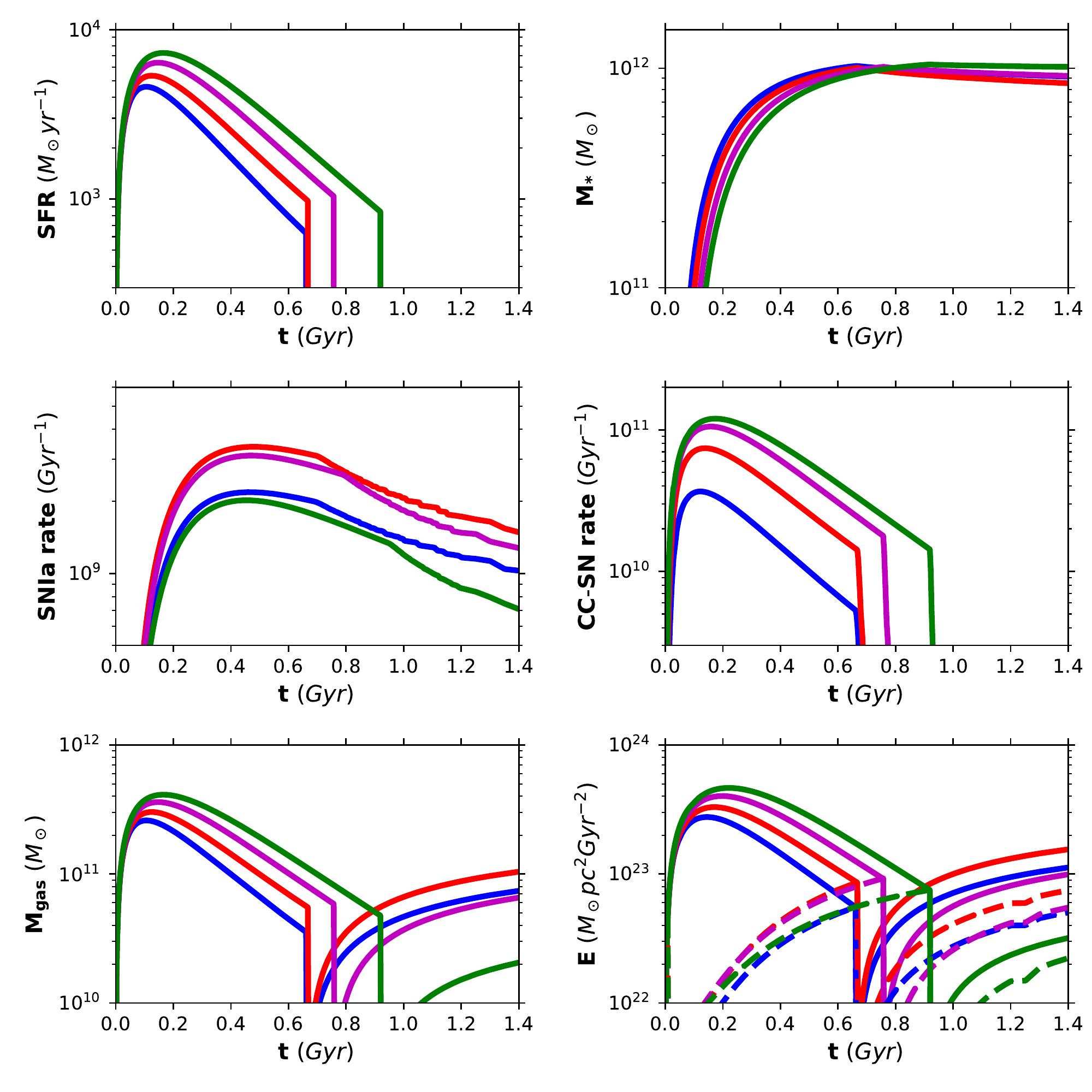}
\caption{Lines are as in Fig. \ref{f:summary10}, but computed for the M1E12 model of  Table \ref{t:models}.} 
\label{f:summary12}
    \end{figure*} 

\subsection{RCSGA 032727-132609}
RCSGA 032727-132609 is a bright lensed galaxy (at the time of its discovery, it was the most luminous lensed galaxy ever known) at redshift $z=1.7037$,
magnified by a factor $\mu=17.2\pm1.4$ by RCS2 032727-132623 galaxy cluster at $z=0.564$. From spectral energy distribution (SED) fitting, the stellar mass was found to be $2\cdot10^{10}M_\odot$ \citep{Wuyts10}.
Different methods give SFR values between $\sim 130 M_\odot yr^{-1}$ \citep{Wuyts10} and $\sim 360 M_\odot yr^{-1}$ \citep{Rigby11}.
As noted in \citet{Rigby11}, the highest SFR value is to be regarded as an upper limit.
The BPT diagram for this object is consistent with no AGN contribution.\\
The abundances in Table \ref{t:abundance_data} are taken from  Keck II/NIRSPEC emission spectrum \citep{Rigby11}.
The $\log$(O/H)+12 values reported in Table \ref{t:abundance_data} refer to the direct, $R_{23}$ and $N_{2}$ methods.
Other abundance indicators (e.g. $N_3O_2$) were used to estimate its metallicity (\citealt{Rigby11}), with derived values which are similar to the ones shown in Table \ref{t:abundance_data}.
 
\subsection{SMACS J0304.3-4402}
This Ly$\alpha$ emitter at redshift $z=1.963$ is magnified by a factor $\mu=42.0\pm8.0$ by a galaxy cluster placed between redshift $0.3$ and $0.5$ \citep{Christensen12a}.
SED fitting reveals a stellar mass of $6.3 \cdot 10^{10}M_\odot$.
The inferred SFR from emission lines lies between $\sim 20$ (friom the $H_\alpha$-detection) and $\sim 90 M_\odot yr^{-1}$ (from the [OII]-detection).
The rest frame UV spectrum shows no AGN contribution.\\
The abundance ratios presented in Table \ref{t:abundance_data} are from the VLT/X-Shooter emission line spectrum by
\citet{Christensen12a,Christensen12b}. O/H have been measured using both direct and $R_{23}$ methods. 

\subsection{SMACS J2031.8-4036}
SMACS J2031.8-4036 is a Ly$\alpha$ emitter at redshift $z=3.51$. It is
magnified by a factor $\mu=15.8\pm7.0$ by a galaxy cluster at
$z=0.331$ \citep{Christensen12a}. The SED fitting reveals a stellar mass
of $ 2.4\cdot10^9 M_\odot$. Its VLT/X-Shooter emission line spectrum indicates no AGN contribution and a SFR between $\sim 15$ and $\sim 30 M_\odot yr^{-1}$, inferred via [OII] and $H_\alpha$ detection, respectively  (\citealt{Christensen12a,Christensen12b}).\\
The abundance data in Table
\ref{t:abundance_data} were drived from the VLT/X-Shooter spectrum. The two
values for $\log$(O/H)+12 are estimated by means of the direct and the $R_{23}$ methods.

\subsection{KBSS - LM1 (composite)}
This composite spectrum 
is the result of the combined analysis of Keck/LRIS and Keck/MOSFIRE observations of a sample of 30 galaxies from the KBSS-MOSFIRE survey \citep{Steidel14}.
Galaxies of this subsample lie in the redshift range $2.113\le z \le 2.572$,
which is optimal to get access to nebular lines as well as integrated OB stars light.
Median values of the stellar mass and the SFR are $\log(M_*/M_\odot)\simeq10.0$ and $\sim50 M_\odot yr^{-1}$, as traced by UV and $H_\alpha$ indicators, respectively.\\
The abundances shown in Table \ref{t:abundance_data} are taken from \citet{Steidel16}.
The three (O/H) values are from different estimators, namely the direct, the $R_{23}$ and the $N_{2}$ methods. 
The direct measurement value is consistent with the one constrained from SED fitting \citep{Steidel16}.

\subsection{Shapley+03 LBG (composite)}
\citet{Shapley03} consider a sample of almost 1000 LBGs at redshift $z\sim 3$ with spectra taken with Keck/LRIS.
In general, the emission line stacked spectrum shows vigorous SF. 
All the four subsamples in which the full sample of galaxies is divided show SFR$>50M_\odot yr^{-1}$.
No evidence of AGN emission is found in the composite spectrum.\\
In Table \ref{t:abundance_data} we show  the abundances derived by  \citet{Shapley03}. 
The O/H abundance is taken from a sample of LBGs originally presented in  \citet{Pettini01}. 

\section{Results}
\label{s:results}

\subsection{The effects of the IGIMF on the galactic star formation history}
\label{ss:physics}
   
In Figures ~\ref{f:summary10}-\ref{f:summary12} we show the impact of the IMF on the evolution
of the SFR, Type Ia and CC-SN rates, as well as gas, stellar mass and 
energetic budget for the starburst models of Table \ref{t:models}. \\

All the models presented in Figures  \ref{f:summary10}, \ref{f:summary11} and \ref{f:summary12} are characterised by SF efficiencies of $5$, $10$ and $20~Gyr^{-1}$, respectively. 
The fact that in each figure the SF efficiency is constant allows us 
to single out the effects of the IMF on the global properties of the galaxy of that particular mass. 

The star formation histories reported in Figures ~\ref{f:summary10}, \ref{f:summary11} and \ref{f:summary12} are strongly dependent on the adopted IMF.\\
In general, the models computed adopting the \citetalias{Weidner11} IGIMF exhibit larger SFR values than the ones computed with a Salpeter IMF. 
This can be explained by the fact that a top-heavy IMF implies larger mass ejection rates from evolved stellar populations, and in particular from massive stars,
with consequently larger gas mass reservoirs at any time,
which also imply larger SFR values.

The differences in the star formation histories caused by the adopted IMF determine also the conditions for the onset of a galactic wind, and consequently the time at which the star formation stops.
We  note that the steepest IGMFs ($\beta=1.6$ and $\beta=2$) produce winds at the earliest times. 
In the case $\beta=1$, in spite of the very high number of CC-SNe (which is roughly proportional to the SFR), the galactic wind occurs later than in the cases with $\beta>1$
because of the larger binding energy of the gas compared to the other models (see Section \ref{ss:chem_evo}).\\
By comparing Figures ~\ref{f:summary10} and \ref{f:summary11}, it is worth noting that a later occurrence of galactic winds in less massive galaxies,
i.e . the downsizing in star formation 
as obtained by \citet{Matteucci94} by means of the the inverse wind model, cannot be reproduced by adopting the IGIMF with $\beta=1$,
and this occurs despite a higher star formation efficiency for the larger mass model. 
It is worth reminding that the inverse wind model can reproduce the observed increase of the integrated [$\alpha$/Fe] with galactic stellar mass in ellipticals (see \citealt{Thomas10}). \\
In \citet{Matteucci94} the winds occuring at earlier times in massive galaxies were obtained by increasing the efficiency of SF with galactic stellar mass,
as we assume here, and with a constant Salpeter IMF. 
The inverse wind effect is visible from the models for the Salpeter IMF as well as for the case $\beta=2$ and $\beta=1.6$.

\begin{figure*}
\centering
\includegraphics[width=.8\textwidth]{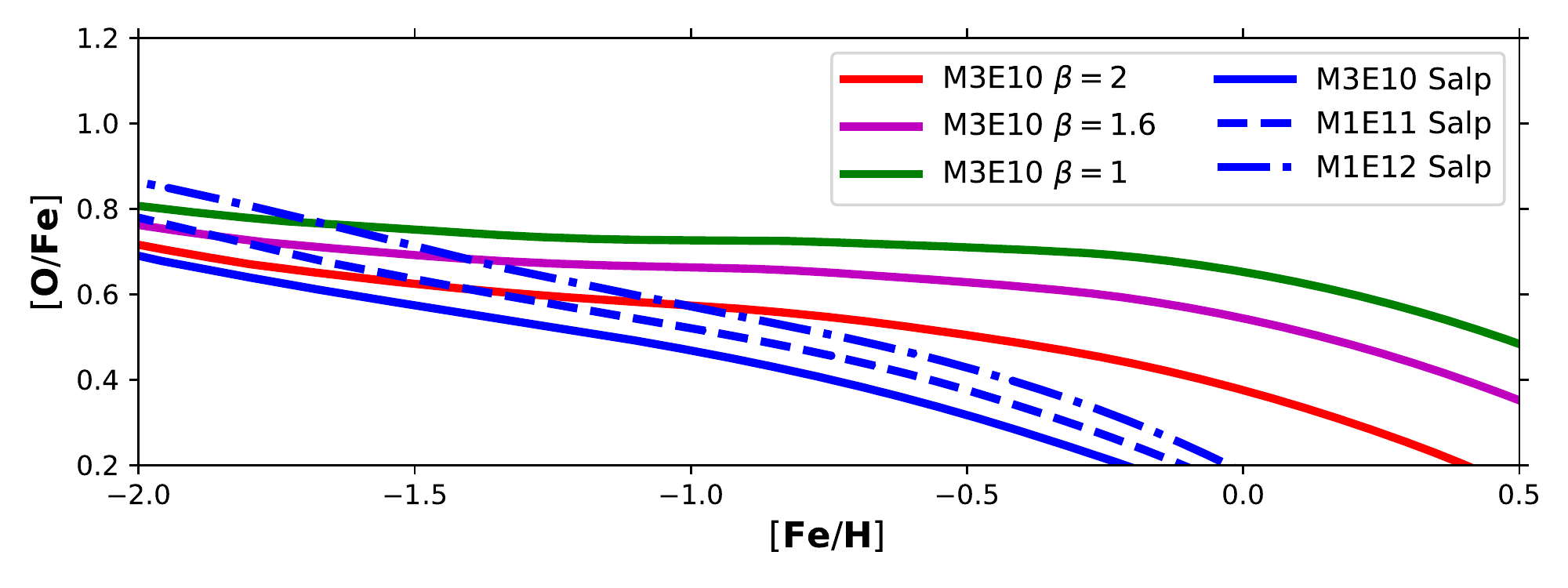}
\caption{Evolution of the interstellar [O/Fe] vs. [Fe/H] relation. Lines are computed for: M3E10 models (solid) with \citet{Salpeter55} IMF (blue) and  \citetalias{Weidner11} IGIMF calculated for $\beta=1$ (green), $\beta=1.6$ (magenta) and $\beta=2$ (red); M1E11 model with \citet{Salpeter55} IMF (dashed blue); M1E12 model with \citet{Salpeter55} IMF (dash-dotted blue).} 
\label{f:OFe_Fet}
\end{figure*}

In Figure \ref{f:summary12} the behaviour of M1E12 models is shown. The results are very similar to what found for M1E11 models.
The same trends shown by the  M1E11 and M1E12 models are explained by the similar IGIMFs, in particular at $\psi>100M_{\odot}yr^{-1}$ (see Figure \ref{f:IGIMF}). The assumption of an upper mass limit for the maximum mass of stellar clusters 
($M_{ecl,max}$) attenuates the dependence of the IGIMF on the SFR at very high $\psi$ values ($\gtrsim 500 M_\odot yr^{-1}$). 

Looking at the lower left panel in Figure \ref{f:summary12}, it is visible that large amounts of gas are restored into the ISM after the onset of the galactic wind. 
The same behaviour is not shown by the lower mass models (Figures \ref{f:summary10} and \ref{f:summary11}), where 
the thermal energy of the gas after the galactic wind is always larger than its binding energy.

Another interesting aspect of Fig. \ref{f:summary12} is that the model with $\beta=1$ gives a Type Ia SN rate which is lower than the one obtained with the Salpeter
IMF, at variance with what shown in Figures  \ref{f:summary10} and, although to a lesser extent, \ref{f:summary11}. 
This is expected from Fig. \ref{f:IGIMF}, where we have seen that the most extreme differences
between the Salpeter IMF and the IGIMF were found when the lowest value for $\beta$ was adopted and at the highest SFR values.
Under such extreme conditions, an IMF remarkably light 
in low- and intermediate-mass stars,  i. e. in the mass range of the progenitors of Type Ia SNe, is possible. 

\subsection{The effects of the IGIMF on chemical abundances}
\label{ss:abundances}

In Figure \ref{f:OFe_Fet}, we show the [O/Fe] vs. [Fe/H] plots computed for the M3E10 models with the IGIMF and with different values of $\beta$ (green, red and magenta lines) as well as for the M3E10, M1E11,
M1E12 models with a \citet{Salpeter55} IMF (blue lines). 

From this Figure we can see that in general, at any time the
\citetalias{Weidner11} IGIMF produces larger [O/Fe] values than the \citet{Salpeter55} IMF.\\
These features can be seen from the different levels of the plateau in the [O/Fe]-[Fe/H] relation
obtained with the M3E10 model assuming different IMFs.
At low metallicity ([Fe/H] $\le -1.5$), higher [O/Fe] values are obtained assuming lower $\beta$ values (i. e. with IMFs heavier in massive stars).
Also the slope of the [O/Fe]-[Fe/H] relation is dependent on the IMF, with a steeper decrease for higher values of $\beta$. \\
A larger extension of the plateau for lower $\beta$ values is again due to a 
larger number of massive stars, which are the first to enrich the ISM with Fe. 
The larger the fraction of massive stars, the higher the metallicity value (as traced by [Fe/H])
at which Type Ia SNe start to contribute significantly to the Fe enrichment,  
thus the higher is the [Fe/H] value for the change in slope of the [O/Fe]-[Fe/H] relation. 
This is a known consequence of the $\lq$time-delay model' \citep{Matteucci12}. 
We also note that in the IGIMF models, a value of [$\alpha$/Fe] larger by 0.2 dex as traced by O extends to much higher metallicity values than with the Salpeter IMF,
reaching values as high as [Fe/H]$\sim$ 0.4. 

The [O/Fe]-[Fe/H] plots for the models M1E11 and M1E12 computed adopting the IGIMF with different $\beta$ values are very similar to Figure \ref{f:OFe_Fet}, hence they are not shown here. 
For comparison, we show only the results of the  M1E11 and M1E12  models obtained for a Salpeter IMF,
to stress that an increase in galactic mass and in SF efficiency leads to an increase of the [O/Fe] value at [Fe/H]=-2, 
without affecting in a substantial way the slope of the [O/Fe]-[Fe/H] relation.  

\subsection{A direct comparison with high redshift starbursts}
\label{ss:abundance_data}

We will now compare the abundance patterns obtained by means of chemical evolution models adopting different IMF prescriptions with the observed abundances of Table \ref{t:abundance_data}.

We will divide the discussion into two parts. First, we will study the behaviour of abundance ratios of volatile elements, namely the elements which are negligibly affected by the presence of dust  (such as N and O). 
Then, we will discuss abundance ratios between refractory elements, i.e. elements severely affected by dust depletion (such as C, Mg, Si, Fe), and compare data with model abundances. 


    \subsubsection{Volatile element ratios}
    \label{sss:volatiles}

\begin{figure}
\centering
\includegraphics[width=1\columnwidth]{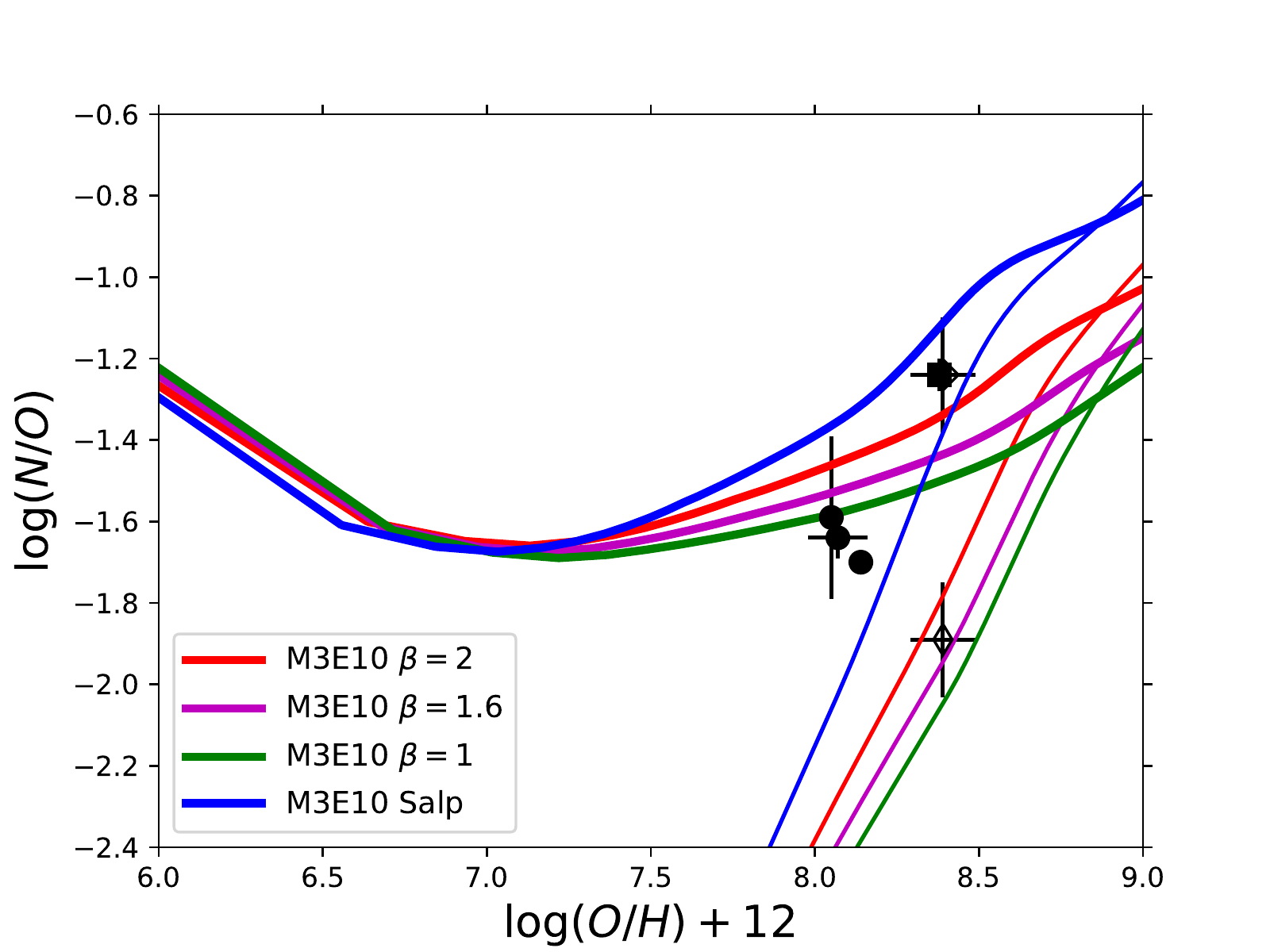}
\includegraphics[width=1\columnwidth]{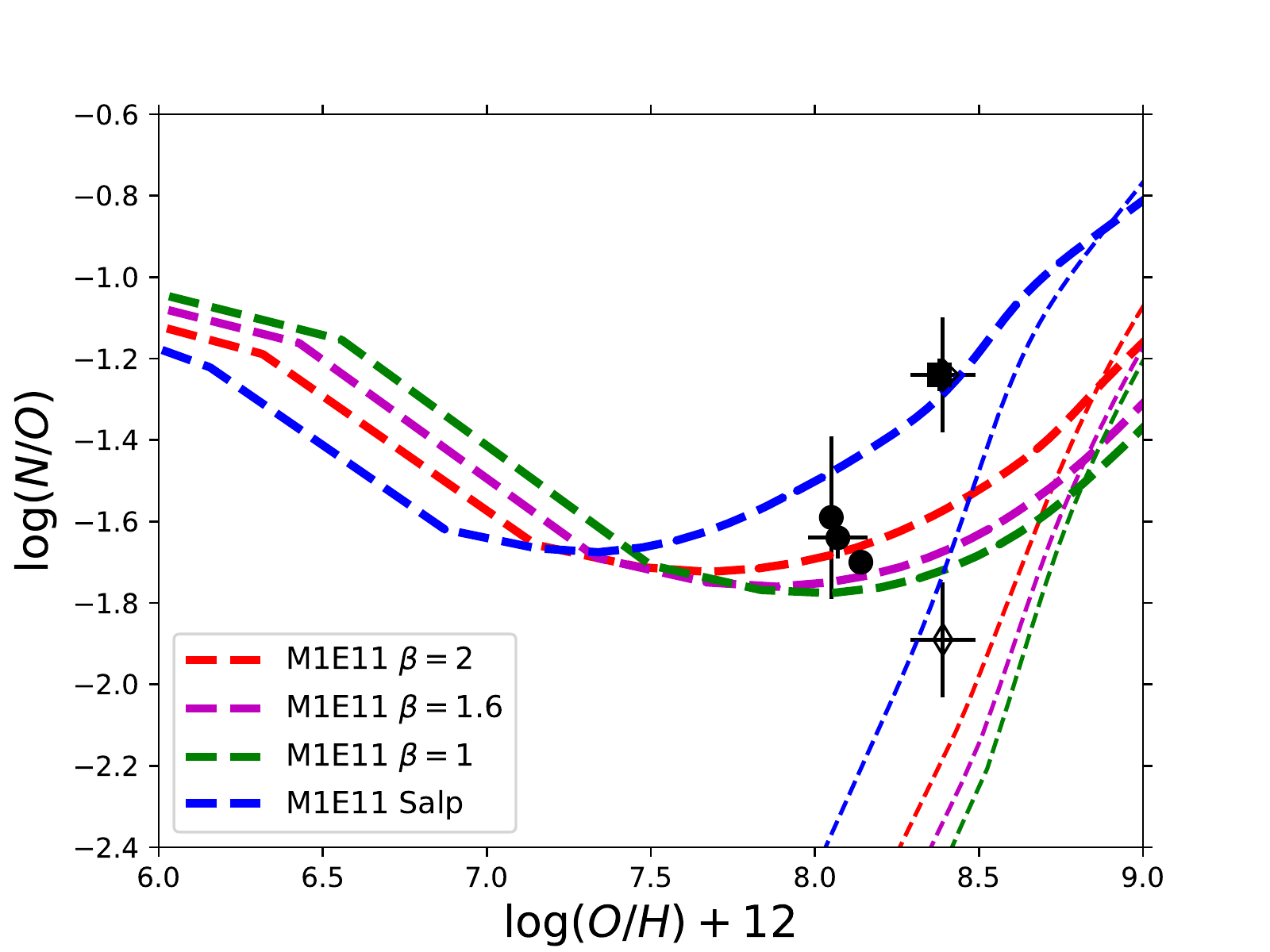}

\caption{$\log$(N/O) vs. $\log$(O/H)+12 adopting \citet{Matteucci86} yields for N (thick lines) and \citet{Meynet02} (thin lines) compared with abundances measured in galaxies of the sample of Table \ref{t:abundance_data}. Upper panel: lines are computed for M3E10 models (solid) with \citet{Salpeter55} IMF (blue) and \citetalias{Weidner11} IGIMF calculated for $\beta=1$ (green), $\beta=1.6$ (magenta) and $\beta=2$ (red). Lower panel: lines are computed for M1E11 models (dashed) with a \citet{Salpeter55} IMF (blue) and \citetalias{Weidner11} IGIMF calculated for $\beta=1$ (green), $\beta=1.6$ (magenta) and $\beta=2$ (red). For both panels: data are from \citet{Steidel16} stacked spectum (filled square); \citet{Rigby11}, \citet{Bayliss14}, \citeauthor{Christensen12a} (\citeyear{Christensen12a}, \citeyear{Christensen12b}) (filled circles); \citet{Pettini02} (thin diamond) and \citet{Teplitz00} (thick diamond).}
\label{f:NO}
\end{figure}    
    
We start our analysis from the (N/O) vs. (O/H) relation, visible in Figure \ref{f:NO}. The analysis of N deserves a special attention, since its origin is still debated. 

For this reason, in the two plots of Figure \ref{f:NO} we show models with different N yields: the \citet{Matteucci86} (thick lines) and the \citet{Meynet02} (thin lines) ones.
The former set assumes that all massive stars produce primary\footnote{Primary production of an element stems directly from the synthesis of H and He. 
In the case of secondary production, the seed for the synthesis must be a heavy element (such as O).}
N, an ad hoc hypothesis that is necessary to explain the high (N/O) ratios observed in low metallicity MW halo stars (\citealt{Israelian04}; \citealt{Spite05}; see also \citealt{Vincenzo18}).\\
The \citet{Meynet02} yields, instead, allow for the production of primary N only in rotating, massive, very low metallicity stars.
In general, this leads to a deficiency of N between low and intermediate metallicities (\citealt{Romano10}; \citealt{Vincenzo16}), at variance with observations.
As for observational data, it is worth stressing that the variation in $\log$(O/H)+12 for a given system due to a different metallicity indicator is typically $\le 0.2$ dex. 
As for the theoretical abundances, here we show only the results obtained for the M3E10 and M1E11 models.

As shown in Figure \ref{f:NO}, in the case of the yields for primary N of \citet{Matteucci86}, 
the adoption of the IGIMF leads to lower (N/O) values at metallicity $\log$(O/H)+12$>7$ with respect to the Salpeter model. 
At lower metallicity, very small variations due to different IMFs are visible in the case of the M3E10 model (upper panel in Figure \ref{f:NO}). \\
The variations between the abundances obtained with the Salpeter IMF and the ones obtained with the IGIMF
increase with metallicity, and in general the lower the $\beta$, the larger the variation.
A maximum variation of $\sim~0.4 $ dex between the (N/O) values with Salpeter and IGIMF is visible at $\log$(O/H)+12$=9$ in the case of $\beta=1$ for the M3E10 model. \\
At all metallicities, larger (N/O) variations between Salpeter and IGIMF are visible in the case of the M1E11 model, with a maximum value of $\sim~0.6$  
at $\log$(O/H)+12$=9$ in the case of $\beta=1$ (lower panel of  Figure \ref{f:NO}).

Similar conclusions can be drawn for the models with the \citet{Meynet02} yields, with a metallicity-dependent increase of the variation between
the (N/O) obtained with the Salpeter and the IGIMF. 

As for the comparison between the abundances observed in the high-$z$ sample of Table \ref{t:abundance_data} and models,
without the primary N yields as suggested by \citet{Matteucci86} it would be impossible to reproduce the (N/O) values measured in
three out of four of the data shown in Fig. \ref{f:NO}. 
This reinforces the results found in previous studies, e. g. on the need of primary N in massive stars to reproduce the (N/O) values measured 
in star-forming galaxies in the Local Universe (e.g. \citealt{Vincenzo16}). 

Figure \ref{f:NO} outlines the role of the adopted IMF in determining the interstellar abundance pattern. In several cases, it is difficult to disentangle between
effects due to the IMF and nucleosynthesis presciptions. 

    
     \begin{figure}
\centering
\includegraphics[width=1\columnwidth]{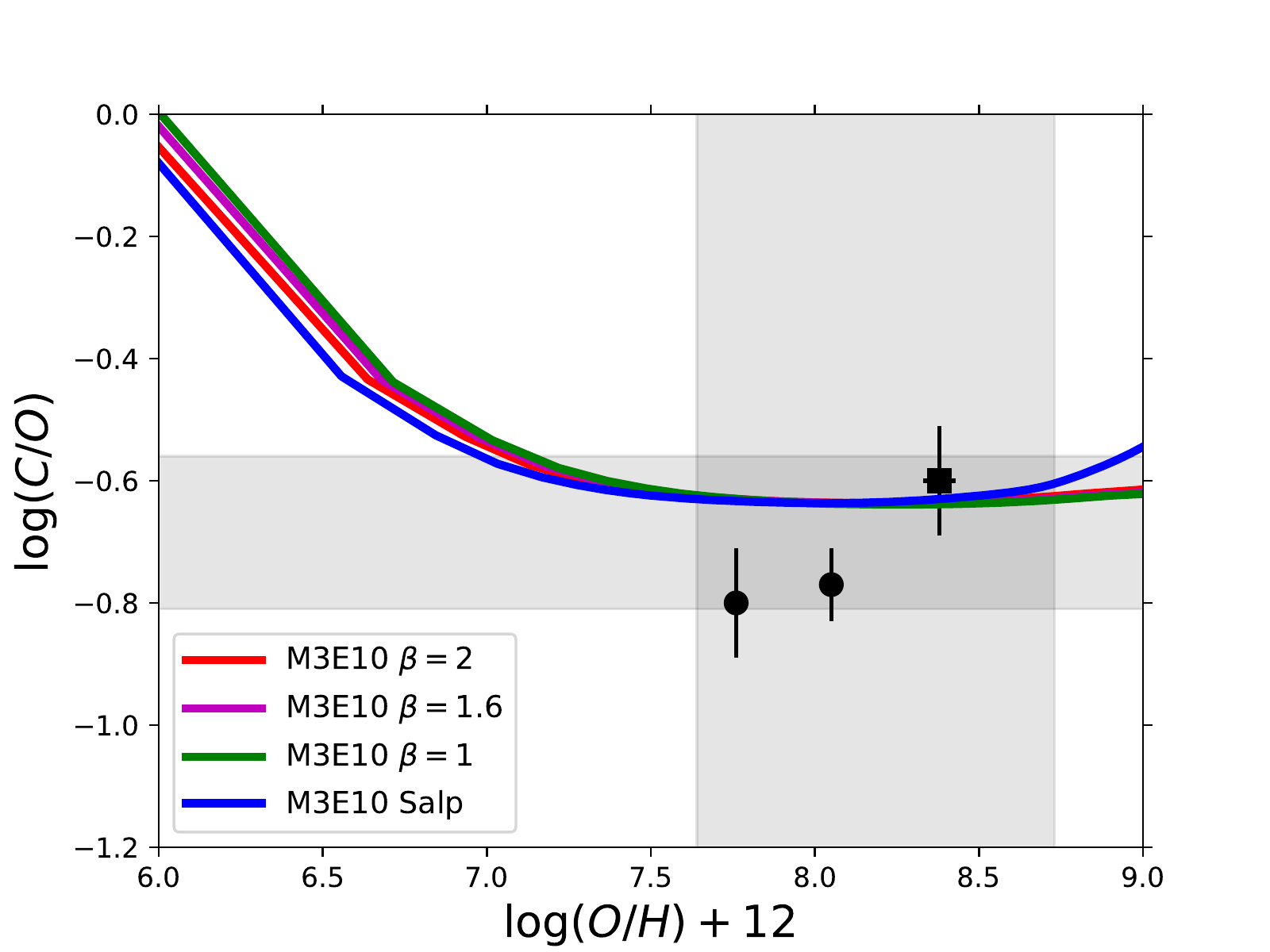}
\includegraphics[width=1\columnwidth]{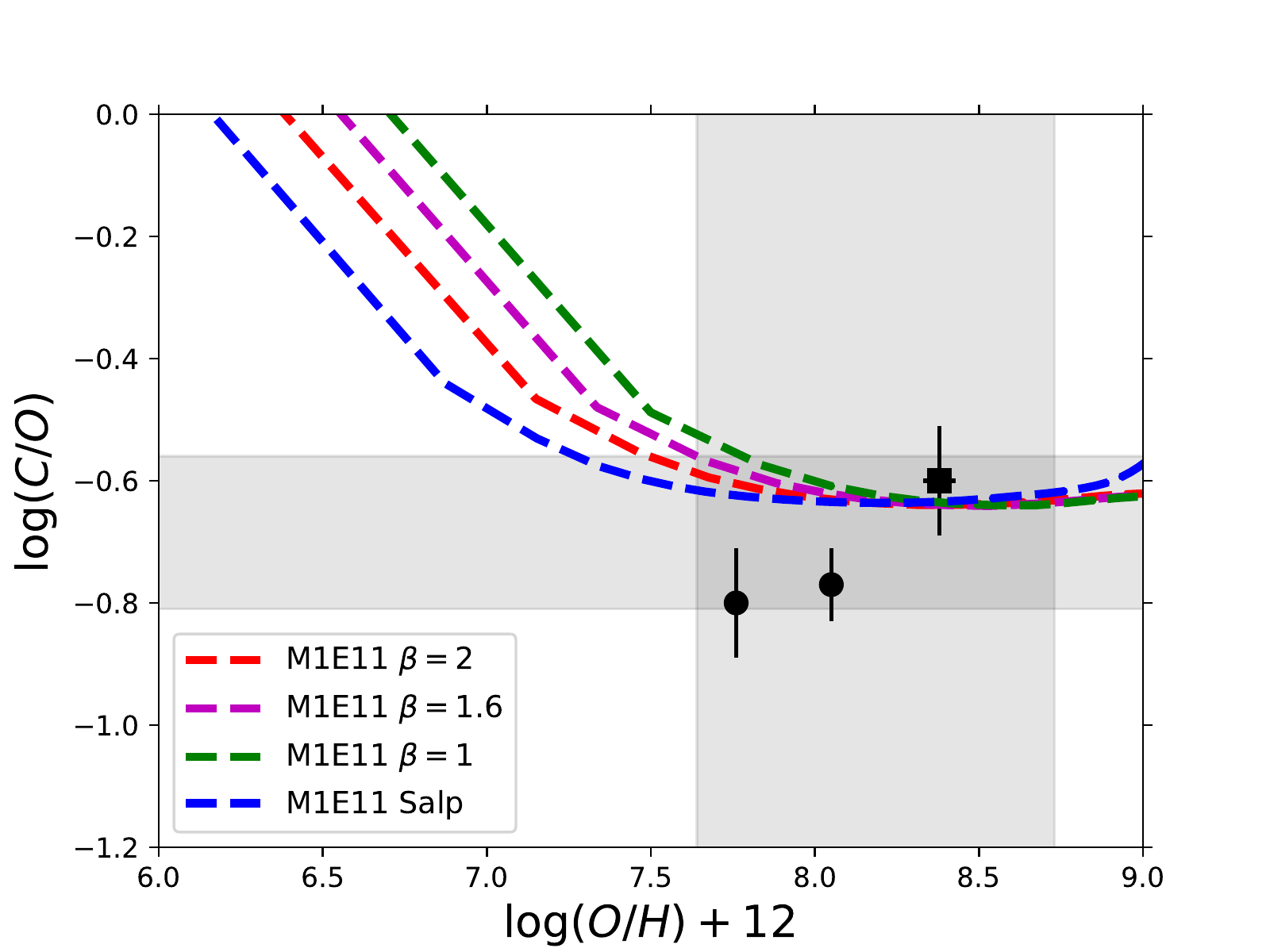}
\caption{$\log$(C/O) vs. $\log$(O/H)+12 computed without taking into account dust depletion and compared with abundances measured in galaxies of the sample of Table \ref{t:abundance_data}.\\
Upper panel: lines are computed for M3E10 models (solid) with \citet{Salpeter55} IMF (blue) and \citetalias{Weidner11} IGIMF calculated for $\beta=1$ (green), $\beta=1.6$ (magenta) and $\beta=2$ (red). Lower panel: lines are computed for M1E11 models (dashed) with \citet{Salpeter55} IMF (blue) and \citetalias{Weidner11} IGIMF calculated for $\beta=1$ (green), $\beta=1.6$ (magenta) and $\beta=2$ (red). For both panels: the shaded regions indicates the $\log$(C/O) confidence region derived from the composite LBG spectrum of \citet{Shapley03} and the $\log$(O/H)+12 characterising the sample of LBGs of \citet{Pettini01}. Other data are from \citet{Steidel16} stacked spectum (filled square), \citet{Bayliss14} and \citeauthor{Christensen12a} ( \citeyear{Christensen12a}, \citeyear{Christensen12b}) (filled circles).}
\label{f:CO}
    \end{figure} 

In summary, the comparison between data and model discussed in this section provides useful suggestions regarding the nucleosynthesis of the volatile elements N and O. However, the analysis of the observational data considered here is not conclusive on whether the IGIMF is to be preferred over the Salpeter IMF to reproduce these particular abundance ratios. 
To this purpose, more insights are provided by the study of the abundance ratios between refractory elements, described in the remainder of this Section. 

\begin{figure}
\centering
\includegraphics[width=1\columnwidth]{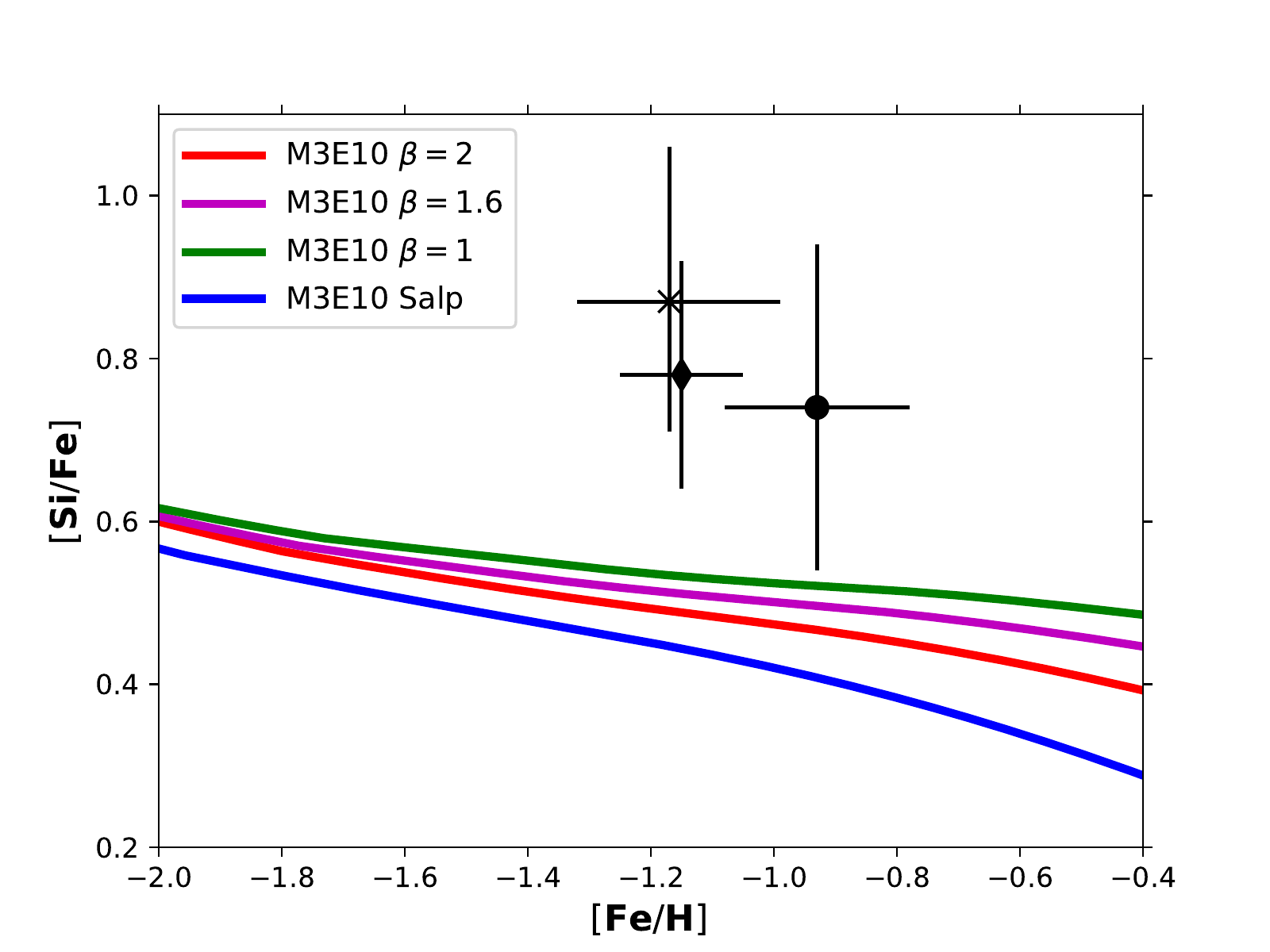}
\includegraphics[width=1\columnwidth]{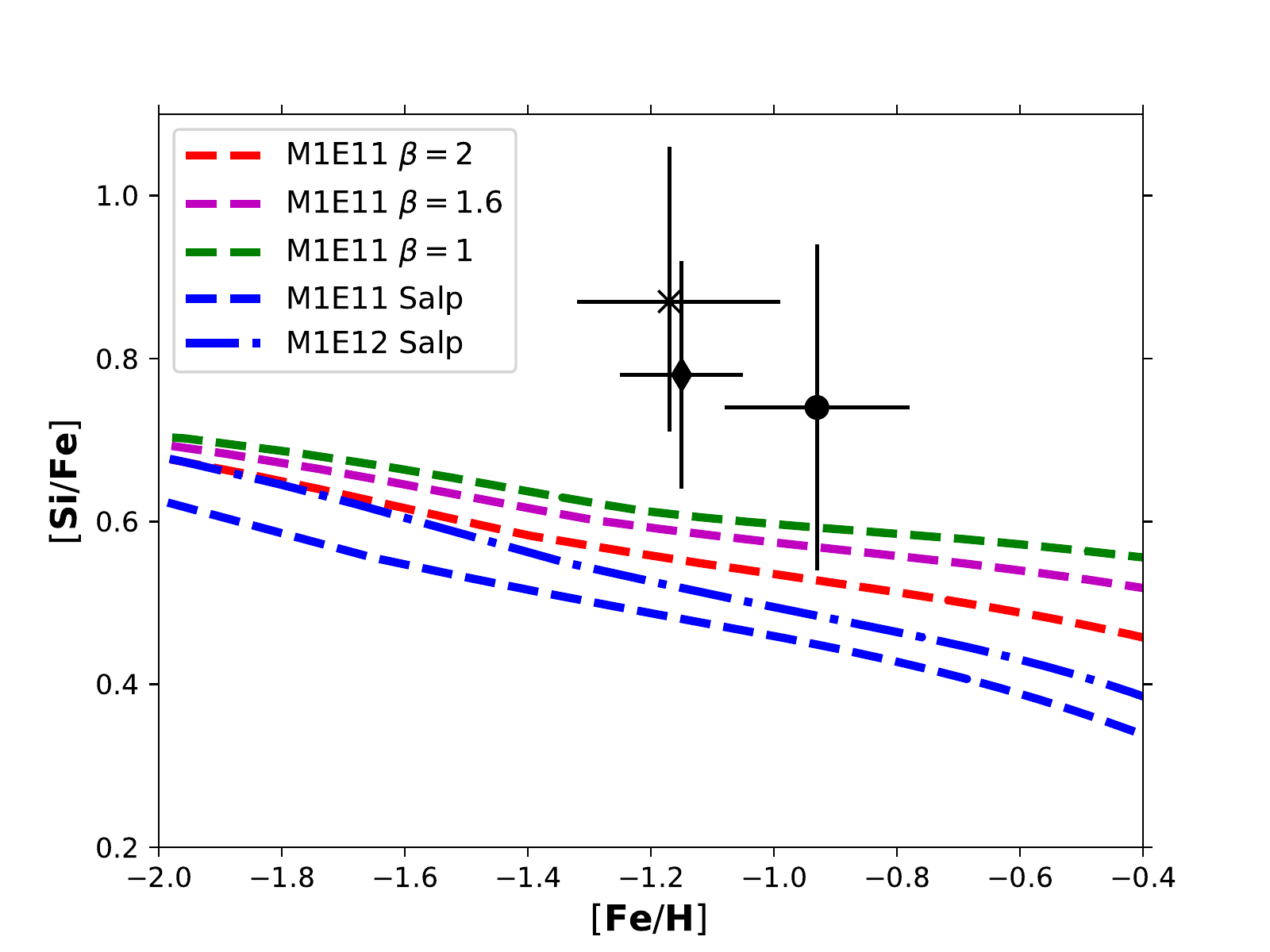}
\caption{[Si/Fe] vs. [Fe/H] computed without taking into account dust depletion and compared with abundances measured in galaxies of the sample of Table \ref{t:abundance_data}.\\
  Upper panel: lines are computed for M3E10 models (solid) with \citet{Salpeter55} IMF (blue) and \citetalias{Weidner11} IGIMF calculated for $\beta=1$ (green), $\beta=1.6$ (magenta) and $\beta=2$ (red). 
  Lower panel: lines are computed for M1E11 models (dashed) with \citet{Salpeter55} IMF (blue), \citetalias{Weidner11} IGIMF calculated for $\beta=1$ (green), $\beta=1.6$ (magenta) and $\beta=2$ (red); M1E12 model with \citet{Salpeter55} (blue dash-dotted).
  For both panels: data are from \citet{Pettini02} (diamond); \citet{Quider09} (cross); \citet{Dessauges10} (filled circle).}
\label{f:SiFe}
    \end{figure}    
    \begin{figure}
\centering
\includegraphics[width=1\columnwidth]{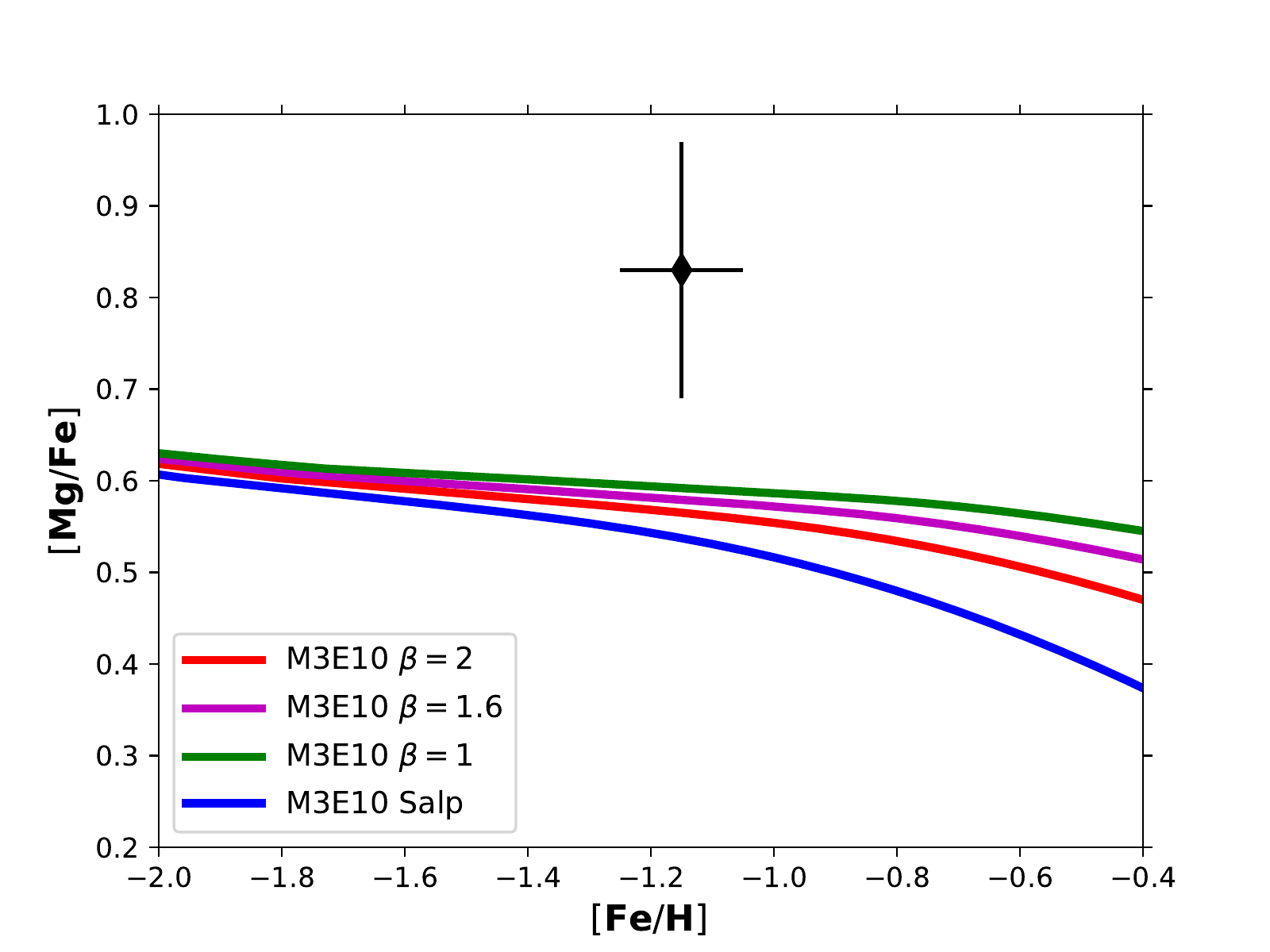}
\includegraphics[width=1\columnwidth]{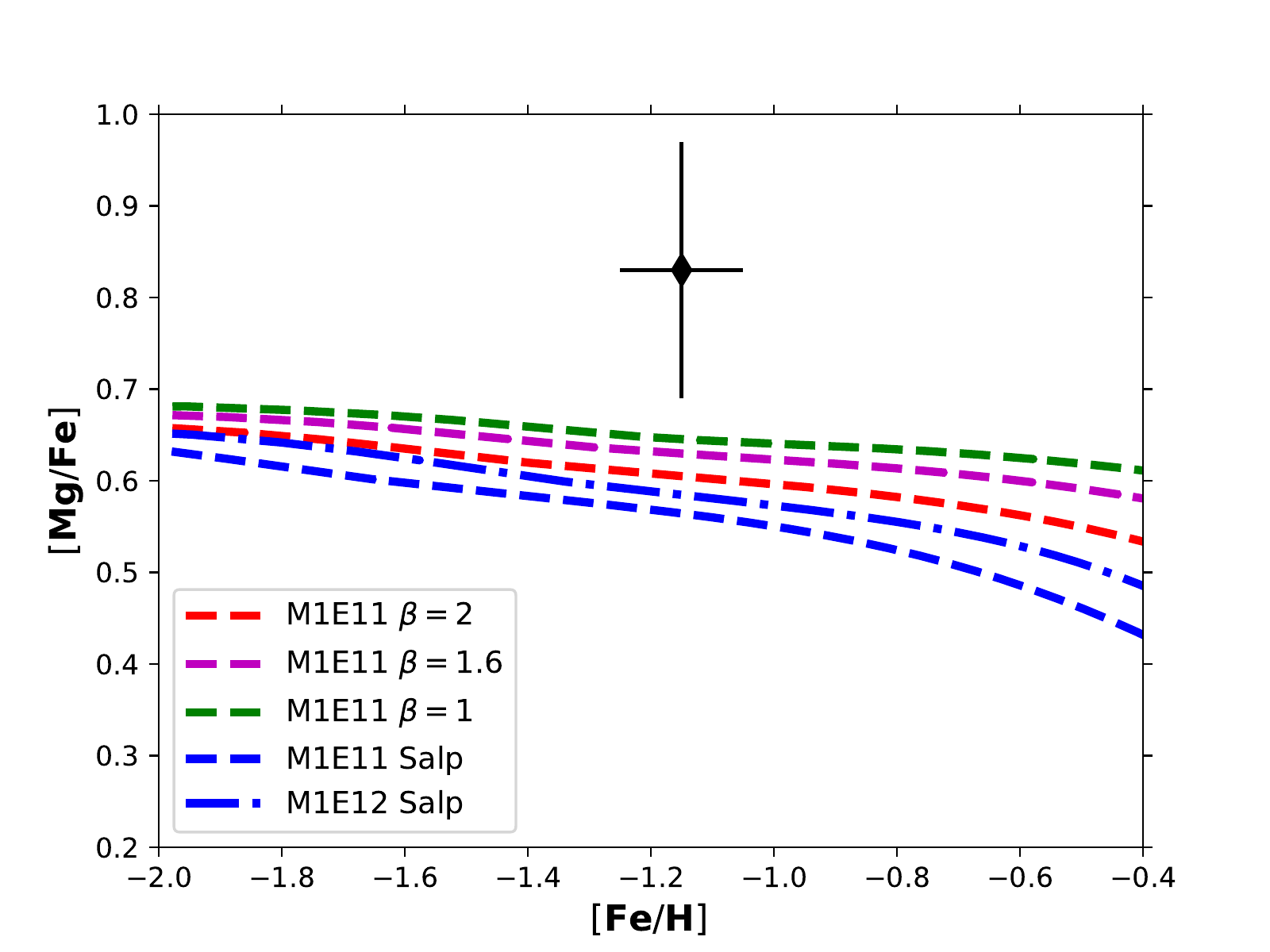}
\caption{[Mg/Fe] vs. [Fe/H] without accounting for dust depletion compared with abundances measured in galaxies of the sample of Table \ref{t:abundance_data}.\\
Upper panel: lines are computed for M3E10 models (solid) with \citet{Salpeter55} IMF (blue) and \citetalias{Weidner11} IGIMF calculated for $\beta=1$ (green), $\beta=1.6$ (magenta) and $\beta=2$ (red).
Lower panel:lines are computed for M1E11 models (dashed) with \citet{Salpeter55} IMF (blue), \citetalias{Weidner11} IGIMF calculated for $\beta=1$ (green), $\beta=1.6$ (magenta) and $\beta=2$ (red); M1E12 model with \citet{Salpeter55} (blue dash-dotted).
 For both panels: data are from \citet{Pettini02} (diamond).
}
\label{f:MgFe}
    \end{figure} 
    
    \begin{figure}
\centering
\includegraphics[width=1\columnwidth]{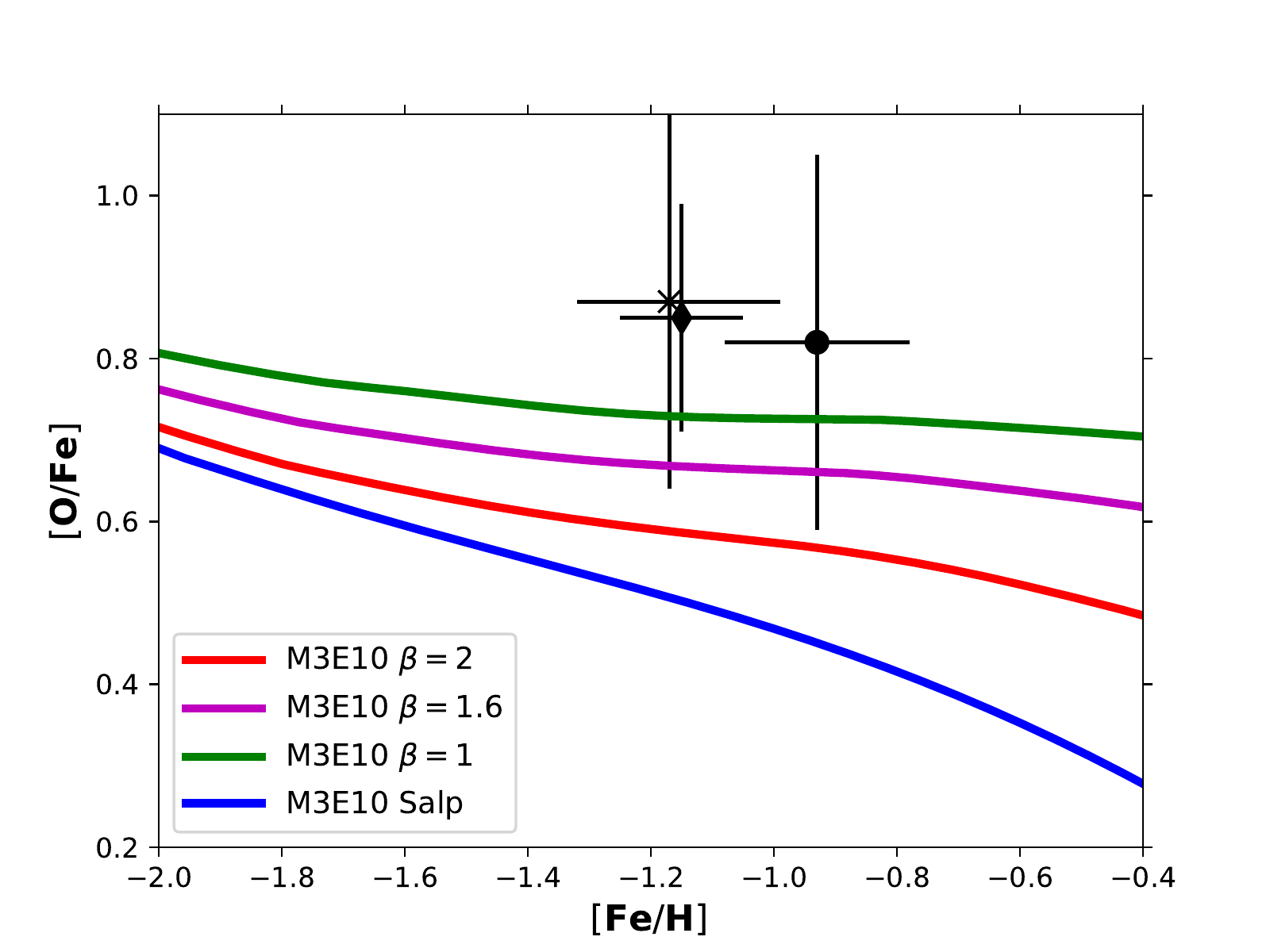}
\includegraphics[width=1\columnwidth]{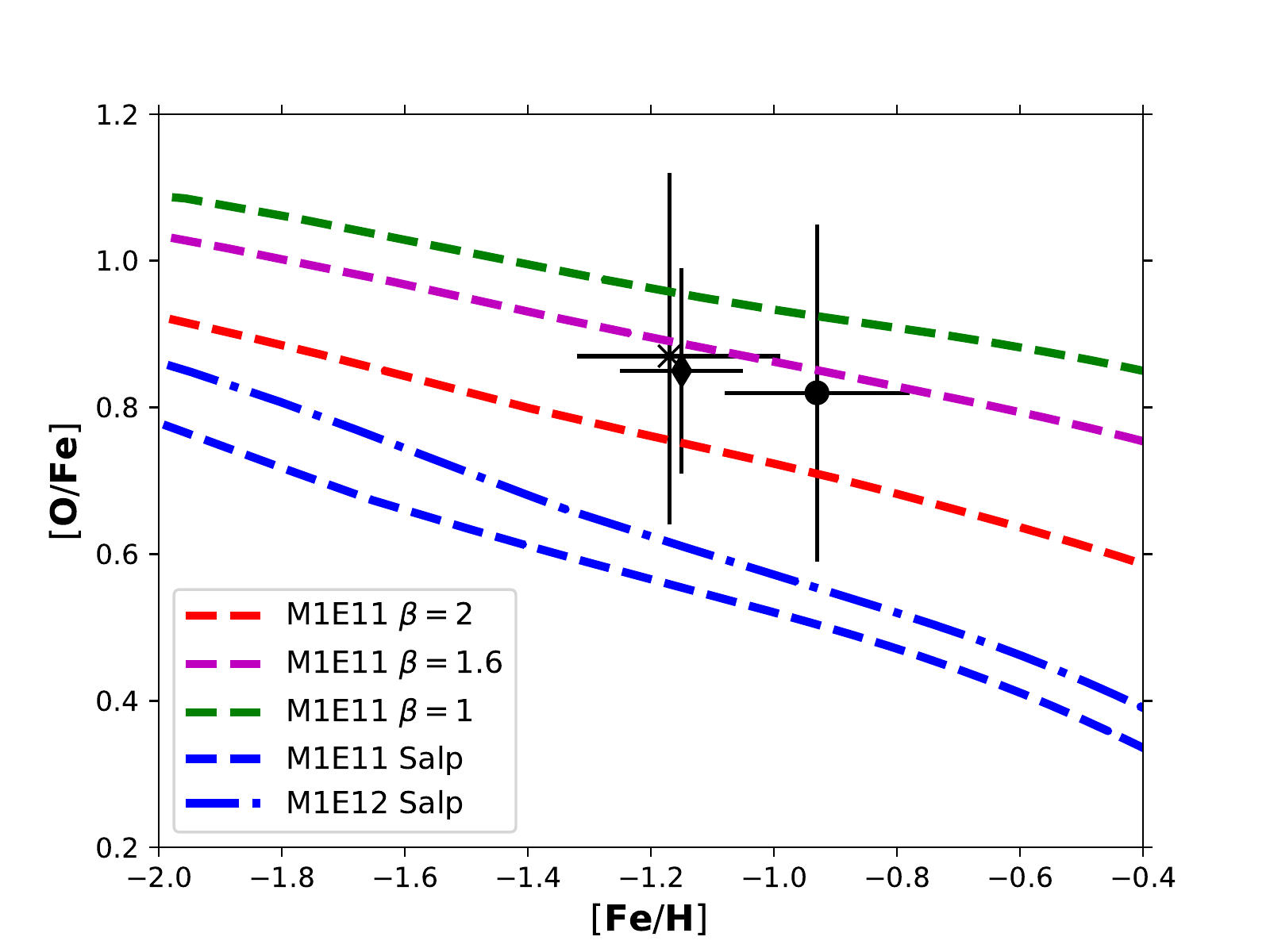}
\caption{[O/Fe] vs. [Fe/H] without accounting for dust depletion compared with abundances measured in galaxies of the sample of Table \ref{t:abundance_data}.\\
Upper panel: lines are computed for M3E10 models (solid) with \citet{Salpeter55} IMF (blue) and \citetalias{Weidner11} IGIMF calculated for $\beta=1$ (green), $\beta=1.6$ (magenta) and $\beta=2$ (red). 
Lower panel:lines are computed for M1E11 models (dashed) with \citet{Salpeter55} IMF (blue), \citetalias{Weidner11} IGIMF calculated for $\beta=1$ (green), $\beta=1.6$ (magenta) and $\beta=2$ (red); M1E12 model with \citet{Salpeter55} (blue dash-dotted).
For both panels: data are from \citet{Pettini02} (diamond); \citet{Quider09} (cross);  \citet{Dessauges10} (filled circle).}
\label{f:OFe}
    \end{figure}

    \subsubsection{Refractory elements abundance ratios}
    \label{sss:refractory}


Figures \ref{f:CO}, \ref{f:SiFe}, \ref{f:MgFe} and \ref{f:OFe} show the calculated (C/O)-(O/H), [Si/Fe]-[Fe/H], [Mg/Fe]-[Fe/H] and [O/Fe]-[Fe/H] relations, respectively,
each one computed without taking into account dust depletion in the ISM 
for M3E10 and M1E11 models, compared to observed abundances from the dataset presented in Sect.~\ref{ss:obs_data} .
We omit for all the plots the results of the M1E12 model with the IGIMF, as they are characterised by SFR values much larger than the ones observed in the systems of our dataset. 

We should highlight that none of the observed abundance ratio involving refractory elements is altered by corrections due to reddening/extinction.\\
In fact, C/O abundances are determined using C3O3 line ratio (\citealt{Garnett95}), which is insensitive to reddening corrections, due to the very similar wavelengths of the lines. On the other hand, other refractory elements are generally studied  by means of absorption line spectroscopy, with  their abundances are just derived from the equivalent width of the lines.

From the (C/O)-(O/H) relation of the M3E10 models (upper panel of Figure \ref{f:CO}) we see that 
varying the IMF does not produce any significant change in the abundance pattern.
In the (C/O)-(O/H) diagram, the effects produced by adopting different IMFs tend to
cancel out, thus producing a very similar behaviour of the abundance ratio vs. metallicity. 
Note that all the models fall in the confidence 
region derived by  stacked spectra by \citet{Shapley03} and the sample of $z\sim3$ LBGs of \citet{Pettini01}. 
Similar conclusions can be drawn from the analysis of the results for the  M1E11 models, with the only difference that
in this case the IGIMF produces higher (C/O) values than Salpeter only at low metallicity (log (O/H)+12 < 7.5), e. g. 
in the case of $\beta=1$ higher by up to a few $0.1$ dex at log (O/H)+12 $\sim$ 6.7, with decreasing differences at increasing metallicity. 
In this case, differences between models with Salpeter and IGIMF are found only outside the observational data range, which does not allow us to prefer any model in particular. Useful information will come later when we will analyse the effects of dust depletion. \\
The (C/O) abundances observed in this dataset confirm the suitability of the yields we are adopting here, which consider rotation in very massive stars ($m>40M_\odot$) at all metallicities (e.g. \citealt{Chiappini03}), also in systems characterised by a star formation history likely much different than the one of the Milky Way.

\begin{figure*}
\includegraphics[width=1.05\textwidth]{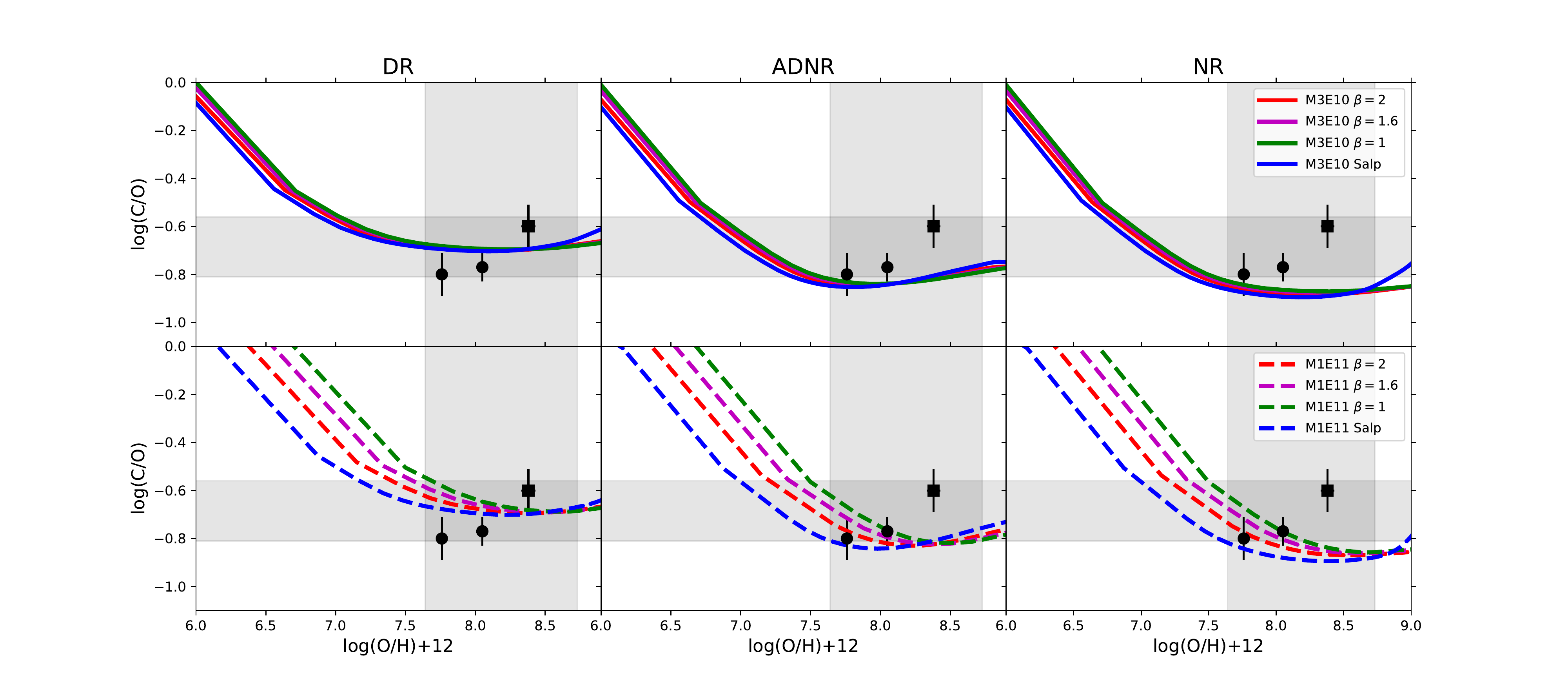}
\caption{Same of Figure \ref{f:CO}, but with models considering dust.}
\label{f:CO_dust}
\end{figure*}

In the [Si/Fe]-[Fe/H] plot for the M3E10 models (Fig. \ref{f:SiFe}, upper panel), for two data points the theoretical abundances lie below the error bars by 0.1 in the best case,
i.e. the models in which the IGIMF is adopted. 
The M3E10 model with the IGIMF and $\beta=1$ lies slightly below the highest-metallicity point.
A marginal overlap between the error bar of the highest metallicity [Si/Fe] value and the M1E11 model with $\beta\le 1.6$ is visible in Fig. \ref{f:SiFe} (bottom panel). 
A sharp disagreement between model results and data is visible also in the [Mg/Fe]-[Fe/H] plots (Fig. \ref{f:MgFe}).  

The observed [O/Fe] values are higher than the ones of the M3E10 models, however the results obtained with the IGIMF and $\beta\le 1.6$ lie within the observational errors (Fig. \ref{f:OFe}, top panel).  \\
The abundances of all M1E11 models are in agreement with the observed [O/Fe] values, whereas the Salpeter IMF produces [O/Fe] values lower than the observed ones,
and also outside of the error bars. Even if the Salpeter M1E12 model produces sligthly overenhanced abundance ratios with respect to the M1E11 one, these are still inconsistent with the data. 
In summary, as for the [$\alpha$/Fe]-[Fe/H] diagrams, the Si and Mg abundances computed with any of our models without taking into account
dust production are lower than the ones observed in high-redshift lensed galaxies, 
whereas the O abundances derived with the models which include the IGIMF are in rather good agreement with the data.

\begin{figure*}
\includegraphics[width=1.05\textwidth]{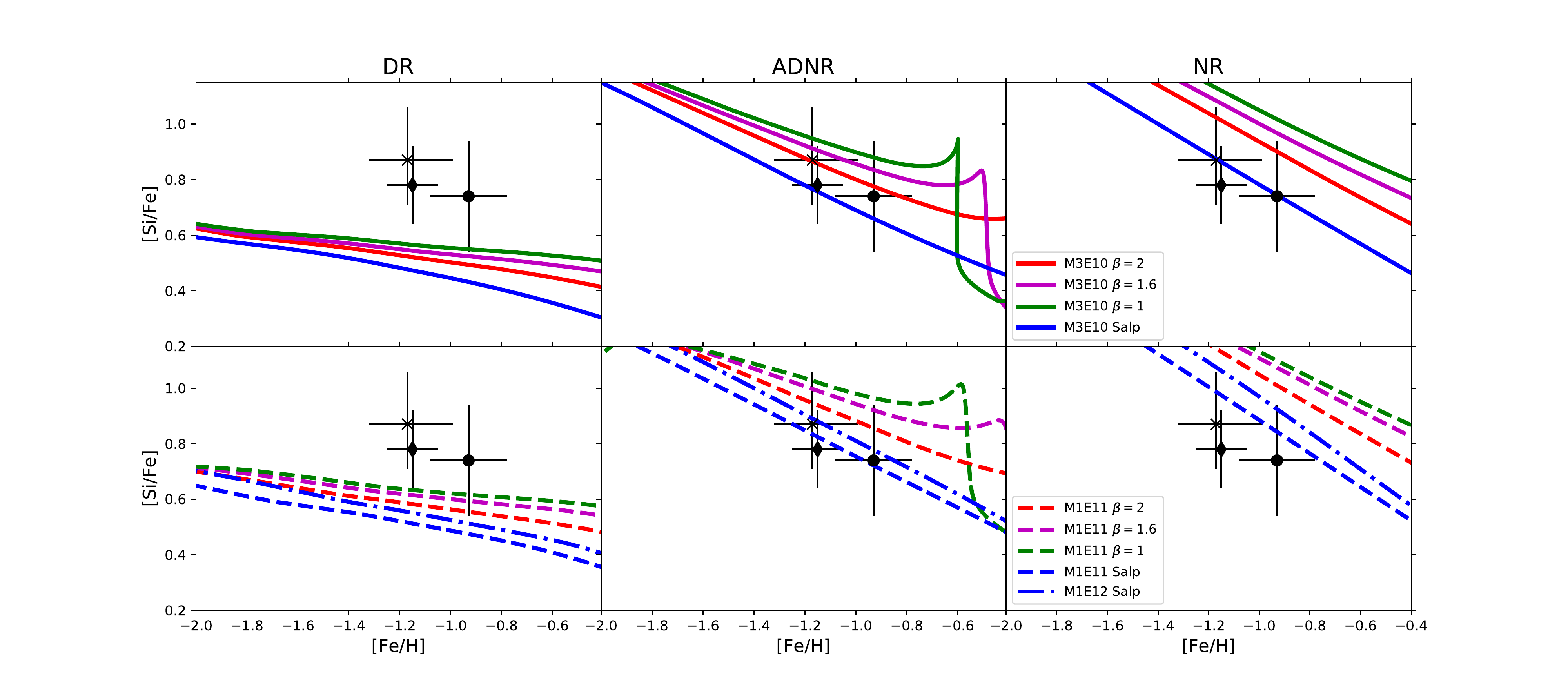}
\caption{Same of Figure \ref{f:SiFe}, but with models considering dust.}
\label{f:SiFe_dust}
\end{figure*}

\begin{figure*}
\includegraphics[width=1.05\textwidth]{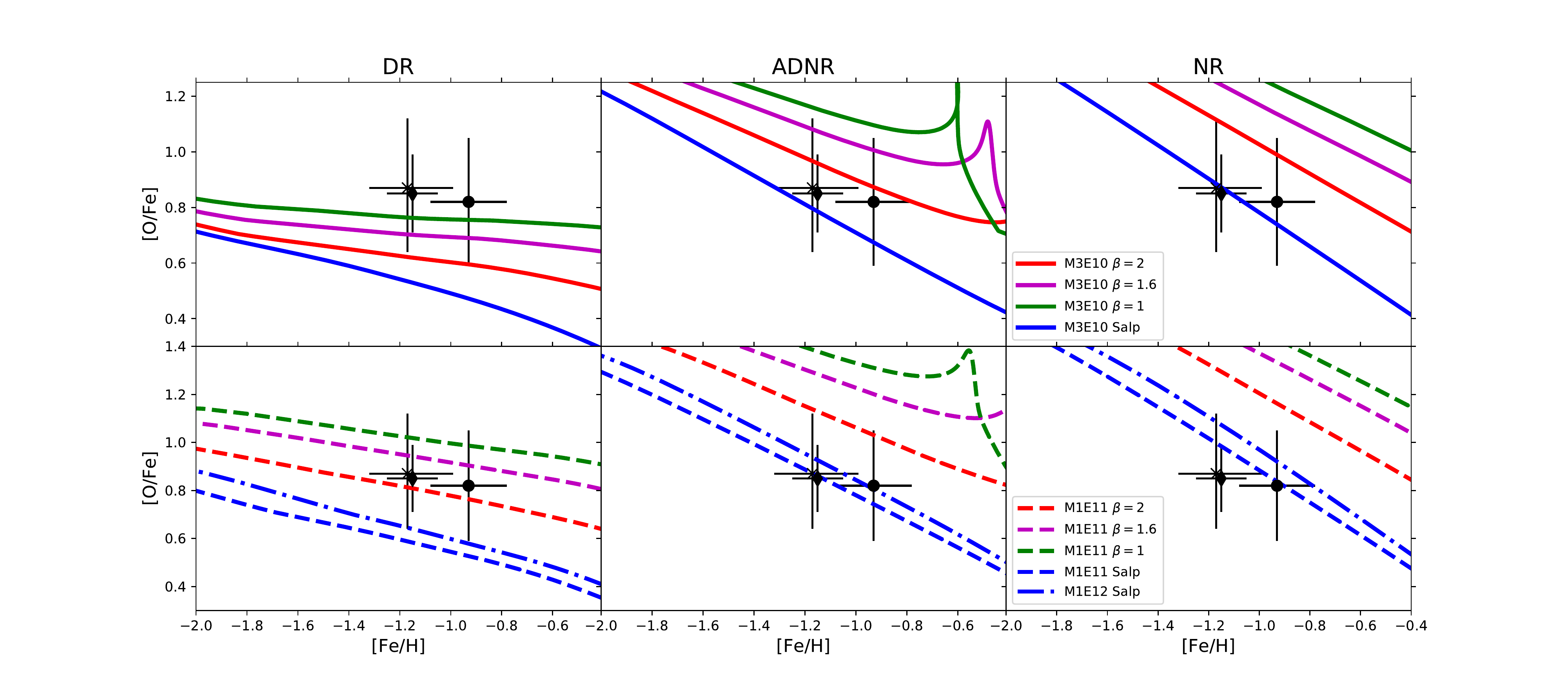}
\caption{Same of Figure \ref{f:OFe}, but with models considering dust.}
\label{f:OFe_dust}
\end{figure*}

In Fig.~\ref{f:CO_dust}, \ref{f:SiFe_dust} and \ref{f:OFe_dust} we show the results of our models computed taking into account the effects of dust depletion for
the (C/O)-(O/H), [Si/Fe]-[Fe/H] and [O/Fe]-[Fe/H] diagrams, respectively, 
compared to the observational abundances derived in high-z galaxies described in Sect.~\ref{ss:obs_data} .
In each figure, the top and bottom rows show the results of the M3E10 and M1E11 model, respectively. 
Such models trace separately the abundances in the gas and dust phases. 
It should also be noted that this is the first time in which the effects of dust are studied in chemical evolution models adopting an IGIMF.\\
We do not show our results including dust for the [Mg/Fe]-[Fe/H] diagram, because for Mg the observed abundances are available only for one system, and also because the theoretical abundances show a behaviour very similar to the one of [Si/Fe]-[Fe/H]. \\
In Fig. ~\ref{f:CO_dust}, \ref{f:SiFe_dust} and \ref{f:OFe_dust} the effects of dust depletion increase as one moves from left to  right. 
In each figure, the leftmost plot shows the results of our $\lq$minimal dust' models, which take into account dust destruction and the reverse shock in SNe dust yields,
and which is therefore dubbed DR, with all the other processes switched off.
The middle plot shows results for models in which dust depletion has intermediate effects and which includes dust condensation, destruction and no reverse shock (ADNR). 
The rightmost plot shows results for our $\lq$maximal dust' models, which includes only dust production from stars with no reverse shock in SNe (NR).

We have tested also a few additional models, not shown in this paper because they show a behaviour similar to the models of Fig. ~\ref{f:CO_dust}- \ref{f:OFe_dust}. 
One model includes dust production in stars with reverse shock in SNe (R) and another has growth, destruction and reverse shock (ADR), in both of which the different prescriptions have small effects in the comparison with data points, relative to the DR model. 
On the other hand, another case in which dust depletion has intermediate effects is the model with destruction but no reverse shock (DNR), which produces results similar to the ADNR model. 

In Fig. ~\ref{f:CO_dust}, the minimal dust DR models show small yet appreciable differences with respect to the ones of Fig. ~\ref{f:CO}.
An evident consequence of including dust depletion is to decrease the (C/O) ratio, as C is refractory and O is not.
For this reason, the (C/O) values of the DR models of Fig. ~\ref{f:CO_dust} are maximum by $\sim 0.1$ lower than the ones of Fig. ~\ref{f:CO}, with a small dependence on metallicity, i. e. with increasing depletion effects at larger values of 12+log(O/H).  \\
The ADNR and NR models show larger C depletion effects and lower (C/O) values. 
The ADNR M3E10 models fall below the dark grey confidence region defined by the range of abundance values from the 
composite LBG spectra (\citealt{Shapley03}), but are still consistent with the values measured in the lensed galaxies SGAS J105039.6+001730 and SMACS J2031.8-4036 (\citealt{Bayliss14,Christensen12a,Christensen12b}). 
As for the M1E11 models, the abundances obtained in most of the ADNR models are by 0.1-0.15 dex lower than the DR models, all still consistent with the composite spectra confidence region.
The results for the NR models are globally similar to the ones obtained in the case of the ADNR models.

As for the M3E10 model, the [Si/Fe]-[Fe/H] plots of Fig. \ref{f:SiFe_dust} show that all the minimal dust models still underestimate the observed abundances.
The models with the IGIMF show higher [Si/Fe] ratios, with distance from the data decreasing with decreasing $\beta$ values, indicating that the abundance pattern of lensed high-z systems might show some signatures of a top-heavy IMF.\\
However, it is plausible that in such systems the effects of dust are important. If this is the case, they are clearly underestimated by our DR models. \\
All the ADNR models account satisfactorily for the observed abundances, in particular the Salpeter and the $\beta=2$ model.
All the NR models in which the IGIMF is adopted overestimate the observed abundances, although the abundances of the $\beta=2$ model are within the $1-\sigma$ error bars of two systems.\\
Similar suggestions come from the analysis of the M1E11 model results. ADNR and DR models show more enhanced [Si/Fe] ratios but still support a Salpeter-like IMF, or suggest that an IGIMF with $\beta<2$ in the observed systems is to be excluded. 

Also in the [O/Fe]-[Fe/H] diagram (Fig. \ref{f:OFe_dust}), 
all the minimal dust M3E10 models underestimate the observed abundances.
Even if all the IGIMF models systematically underproduce the observations, the $\beta<2$ IGIMF models yield abundance values within
the error bars. On the other hand, all the M1E11 IGIMF models show abundance ratios consistent with the observations.
In both cases, the models with the Salpeter IMF underestimate the observed abundance pattern. \\
In all the other models where the effects of dust are more significant, the analysis of the M3E10 models support a Salpeter IMF
or a moderate top-heavy (IGIMF with $\beta<2$) in high-redshift galaxies, whereas all the M1E11 models with dust exclude the IGIMF.

\begin{figure*}
\includegraphics[width=1.05\textwidth]{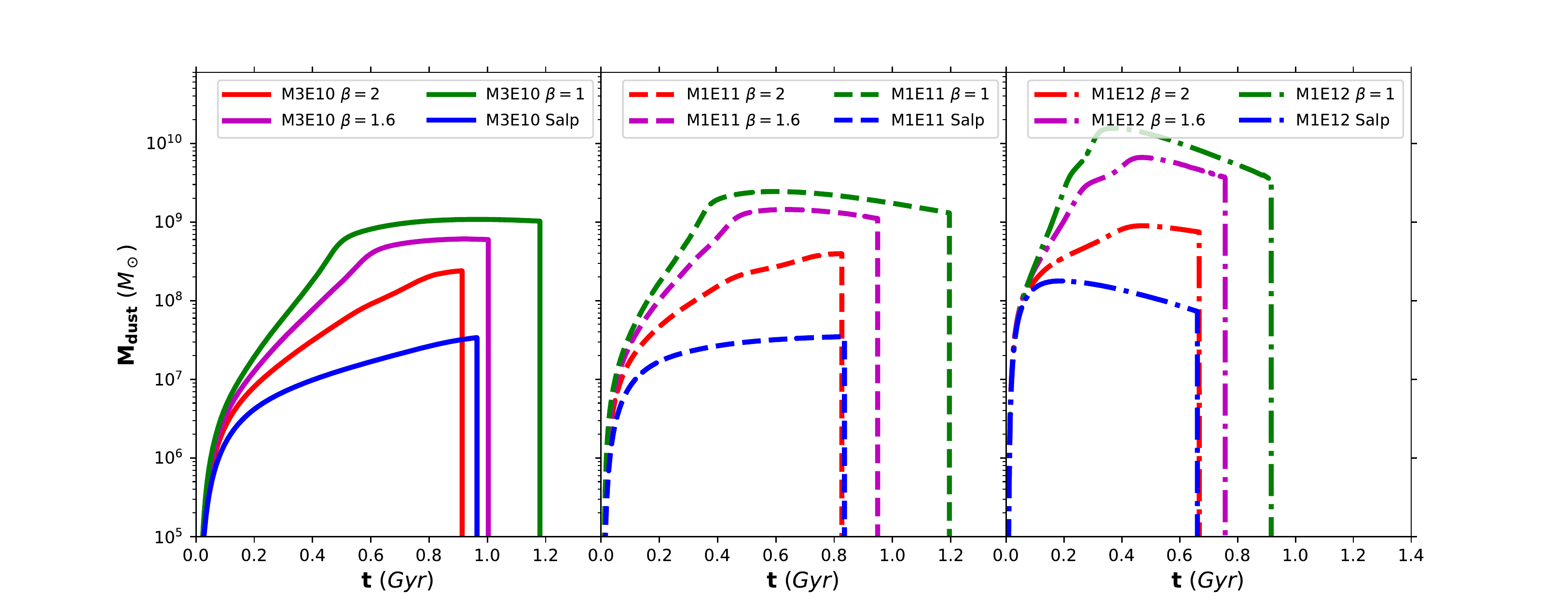}
\caption{ Dust mass as a function of time for our models computed with a \citet{Salpeter55} IMF (blue lines) and with \citetalias{Weidner11} IGIMF with $\beta=1$ (green lines), $\beta=1.6$ (magenta lines) and $\beta=2$ (red lines). In the left, middle and right panel we show our results computed for the M3E10 (solid lines), M1E11 (dashed lines) and M1E12 (dash-dotted lines) models, respectively.}
\label{f:Mdust}
\end{figure*}

\subsection{Discussion}
Previous chemical evolution studies including dust (\citealt{Pipino11}) failed in reproducing the abundances of cB58 adopting a \citet{Salpeter55} IMF,
but they did not considered differential depletion, i.e. different elements depleted in dust in different proportions.
In this work instead, the use of elemental dust yields dependent on mass and metallicity of the stars allows us to account for differential dust depletion.
If one focuses on a single mass model (M3E10 or ME11), our study shows that the effects of dust may produce variations in the abundance ratios larger than the ones driven by a different IMF. 

If the effects of dust are marginal in our sample of
high-redshift star-forming galaxies, our IGIMF models account for the observed abundance pattern better than the ones with a Salpeter IMF.
On the other hand, if the effects of dust are moderate or high, the models with a Salpeter IMF or with the IGIMF with $\beta=2$ allows us to reproduce the data,
whereas the ones with more extreme assumptions for the IGIMF (i.e. $\beta\le1.6$) do not. 

Our study is the first in which an extended database of measured abundances from different chemical elements in lensed galaxies is compared with results from chemical evolution models. 
More observations are certainly needed in the future to shed more light on how the combined effects of dust depletion and IMF determine the abundance pattern of high-redshift galaxies. 

In Fig. \ref{f:Mdust} we show the time evolution of the dust mass for our models with a Salpeter IMF and with an IGIMF, all computed adopting dust prescriptions as in the ADNR model (i. e. the model with intermediate dust depletion effects in Fig.~\ref{f:CO_dust}-~\ref{f:OFe_dust}).  
Fig. \ref{f:Mdust} is useful to assess the timescale for the the buildup of dust and what is the role of the IMF in this process.
In all models, a progressively steeper increase of the dust mass as a function of time is found as the value of $\beta$ decreases.
This highlights directly not only that a more top-heavy IMF produces a larger dust content, but also a faster buildup. 
As discussed in previous works (\citealt{Mattsson11}, \citeyear{Matsson15}; \citealt{Gall11}; \citealt{Gioannini17}), a fast growth of dust goes in lockstep with a rapid buildup of refractory elements, clearly strongly dependent on the SFH. 
In starburst galaxies the buildup of the metals occurs on a particularly rapid timescale, with a supersolar metallicity reached already at $\sim$0.1 Gyr (e.g., \citealt{Calura14}).\\ 
In the case of a Salpeter IMF, in the M1E11 and M1E12 models 
the bulk of dust mass is already present at $\sim$0.25 Gyr after the beginning of star formation. 
Clearly, the rapid buildup of the dust also depends on the infall timescale, which is shorter in larger systems.  
With these particular prescriptions for dust, the dust mass values attained with a Salpeter IMF after a few 0.1 Gyr are generally between $\sim~10^7$ and $\sim~10^8~M_{\odot}$, lower than the values observed in a large fraction of galaxies in the Herschel sample at comparable redshift, which in many cases show larger $>10^8~M_{\odot}$ 
(\citealt{Calura17}, \citealt{Pozzi20}). \\
The same is not true for all the models with an IGIMF, which present dust mass values in excess of  $10^8~M_{\odot}$. This is particularly evident for models with $\beta=1$, which reach such values already after 0.2 Gyr.\\
This latter result is apparently in contrast with what found from abundance patterns. However, uncertainties in the abundance data and in the elemental dust yields (e.g. \citealt{Gall18} and references therein) have to be considered. Furthermore, in general, models with a moderate top-heavy IGIMF (i.e., $\beta=2$) reasonably satisfy both the abundance and dust mass constraints.\\[0.1cm] 

The necessity to adopt a top-heavy IMF to solve the ``dust-budget  crisis'' (\citealt{Rowlands14}) was already discussed by other authors (\citealt{Mattsson11}; \citealt{Gall11}; \citealt{Valiante14} and references therein). 
However, this aspect was never highlighted before in the context of the IGIMF. 
Another important aspect of our study is that the interplay between varius parameters, including the IMF, in determining the abundance pattern in high-redshift galaxies and how these parameters influence the growth of the dust mass.
At present, infrared-based determinations of the dust mass in Lyman-Break galaxies or in the sample considered in this
work is lacking, but it will be valuable in the future to better single out the role of such parameters in determining
the properties of high-redshift starbursts. 

\section{Conclusions}
\label{s:conclusion}
In this work, we have studied the effects of the  integrated galactic IMF (as defined in \citetalias{Weidner11}) as well as interstellar dust evolution on the chemical evolution of high redshift starburst galaxies.

The IGIMF has a strong impact on the star formation history and chemical evolution of galaxies.
In the IGIMF theory, in a galaxy the maximum mass of a stellar cluster increases with the SFR. 
In general, the higher is the SFR value, the larger is the fraction of massive stars (and the flatter is the IGIMF). \\
This affects several basic properties of galaxies, such as mass return from stellar populations, supernova rates and even the star formation history; we have shown how the adoption of an IGIMF produces remarkable differences in these properties with respect to a classical \cite{Salpeter55} IMF.\\
Within the IGIMF theory, one key, yet unkown parameter which regulates the number of massive stars is the slope $\beta$ of the embedded cluster mass function; in
this work, we have tested three different values for this quantity. 

In principle, the stellar populations of 
systems characterised by an intense star formation activity and by strong SFR values might present an overabundance of massive stars, 
whose signature might be present in the interstellar abundance pattern. 

In order to probe the IMF of starburst galaxies, we have compared the abundance patterns computed by our models
with the abundances observed in spectra of high redshift starburst galaxies, mainly LBGs and Lyman-$\alpha$ emitters.\\
In our models, we have also taken into account the effects of dust depletion, which can have a strong impact on interstellar abundances,
and which is fundamental to interpret the observed abundance pattern. 

Our results can be summarised as follows.
\begin{enumerate}
\item In all our models, the adoption of the IGIMF causes the increase of the rate of star formation  with respect to the Salpeter IMF. 
In particular, the highest SFRs are obtained with the lowest values for the slope of the embedded cluster mass function $\beta$. 
This is a consequence of the behaviour of the IGIMF: the lower is $\beta$, the more top-heavy is the IMF.
A high percentage of massive stars as due to a top-heavy IMF cause large quantities of gas to be restored into the ISM with  CC-SN explosions, which in turn favour higher star formation rate values (as SFR $\propto$ M$_{gas}$). 
On the other hand, we find a longer duration of the star formation phase for lower $\beta$ values ($\beta=1$ in particular), due to later galactic winds which, in our picture, cause the end of star formation. \\
In general, the higher the $\beta$ value, the smaller the differences in the models with respect to the Salpeter IMF. 
In the models with $\beta=2$, the  occurrence of the galactic wind is comparable to the one obtained with a Salpeter IMF.\\
For the most extreme assumptions of the $\beta$ value ($\beta=1$), we have found a reduced Type Ia SN rate with respect to the Salpeter in M1E11 and M1E12 models.
Moreover, we have found that with $\beta=1$ the downsizing in star formation, i.e. winds occurring at earlier times in more massive galaxies, is not reproduced in the M3E10 and M1E11 models, despite a higher star formation efficiency for the larger mass model is adopted. 

\item The different star formation histories obtained with different IMFs have an impact on the evolution of chemical abundances.
For a given galaxy mass, the more top-heavy the IMF (i.e. the lower the value for $\beta$), the faster is the growth of the metal content.  
As for the [$\alpha$/Fe]-[Fe/H] diagram, in general,  the lower the $\beta$,
the higher the overabundance of $\alpha$-elements, and the higher the metallicity at which the [$\alpha$/Fe] starts to deviate from
the initial plateau.\\
Our study also highlights the interplay between the IMF and the star formation efficiency in defining 
the insterstellar abundance pattern. In particular,  
when the IGIMF is adopted, the lowest mass model can be characterised by values for the $\alpha$-enhancement
larger than what obtained with the Salpeter IMF in the most massive galaxy, in which a much larger star formation efficiency is adopted. 

\item We have collected a dataset of chemical abundances measured in high-redshift, star-forming galaxies. The sample consists of high-quality abundances for several elements (C, N, O, Mg, Si, Fe) as measured in lensed galaxies at $2 \lesssim z \lesssim 3$ and two $\lq$composite' spectra from sizable sets of high-z systems, obtained by means of observations in non-lensed fields.
The observational results have been compared with our model results. \\
Some of the measured abundance ratios between volatile elements (O, N) are in agreement with our results in which the IGIMF is adopted, in particular in the the (N/O) vs. (O/H) plots.
However, a large scatter in the data and large uncertainties in the stellar yields, in particular regarding N production,
makes difficult the interpetation of the abundance ratios and the degeneracy between the effects of the IMF
and the ones of stellar nucleosynthesis prevents us from reaching any firm conclusion.

\item To interpret the abundances for refractory elements, we have used models which can account for differential
  dust depletion (i.e. different elements depleted into dust in  different proportions), and we have tested various assumptions regarding the processes that regulate the evolution of dust grains.
All the models with minimal dust production (i.e. where dust destruction and the reverse shock in SNe are considered, whereas dust growth is not) underestimate the observed pattern in the [Si/Fe]-[Fe/H], [Mg/Fe]-[Fe/H] and [O/Fe]-[Fe/H] diagrams.
Under these conditions, the models with the IGIMF give [$\alpha$/Fe] values in better agreement with the observations with respect to the ones with a Salpeter IMF.\\
On the other hand, if the effects of dust are important, our analysis suggest a Salpeter or an IGIMF with $\beta\geq2$, i.e. slightly more top heavy than the Salpeter but not particularly extreme.
Several previous studies have evidenced that high-redshift star-forming galaxies can contain large amounts of dust
(e. g., \citealt{Calura17} and references therein,  \citealt{Gall18}). In the light of this, a scenario in which the effects of dust are negligible seems not plausible.\\
We have also calculated the evolution of the dust mass with time for galaxies of different stellar mass and IMF, showing that an at least moderate top-heavy IGIMF is required to solve the "dust budget crisis" problem (e.g. \citealt{Valiante14}).\\

Other works support a top-heavy IMF in high-redshift starbursts, in particular a few studies based on the interpretation by means of chemical evolution models of the abundances of rare isotopes such as $^{13}$C, $^{15}$N, $^{17}$O and $^{18}$O ratios, detected in the infrared band \citep{Romano17,Zhang18}.
In the future, it will be important to take into account the IGIMF in models to interpret such measures,
and possibly assess better the role of downsizing in the star formation histories of such systems. 

On the observational side, the Multi Unit Spectroscopic Explorer (MUSE) mounted on the VLT has allowed the identification and the spectroscopic confirmation of hundreds of multiple images in the redshift range $2\le z \lesssim 7$ (e. g., \citealt{Vanzella17} and references therein).\\
This has also enabled the production of highly accurate lens models, useful to determine and interpret absolute physical quantities such as luminosities, stellar masses, star formation rate values and abundances of high-redshift galaxies (e. g., \citealt{Meneghetti17}).
In the future, it will be important to perform high-resolution spectroscopic follow-ups of the most magnified 
sources, in order to derive precise abundance ratios and to extend the observational sample presented in this work.

\end{enumerate}

\section*{Acknowledgements}
M. P., F. M. acknowledge financial support from the University of Trieste (FRA2016).
F. C. acknowledges the support from grant PRIN MIUR 2017 - 20173ML3WW\_001 and the INAF main-stream (1.05.01.86.31).
F.V. acknowledges the support of a Fellowship from the Center for Cosmology and AstroParticle Physics at The Ohio State University. 
The authors thank the anonymous referee for careful reading of the manuscript and useful suggestions.
\bibliography{IGIMF_starbursts_revised2}

\begin{thebibliography}{}
\makeatletter
\relax
\def\mn@urlcharsother{\let\do\@makeother \do\$\do\&\do\#\do\^\do\_\do\%\do\~}
\def\mn@doi{\begingroup\mn@urlcharsother \@ifnextchar [ {\mn@doi@}
  {\mn@doi@[]}}
\def\mn@doi@[#1]#2{\def\@tempa{#1}\ifx\@tempa\@empty \href
  {http://dx.doi.org/#2} {doi:#2}\else \href {http://dx.doi.org/#2} {#1}\fi
  \endgroup}
\def\mn@eprint#1#2{\mn@eprint@#1:#2::\@nil}
\def\mn@eprint@arXiv#1{\href {http://arxiv.org/abs/#1} {{\tt arXiv:#1}}}
\def\mn@eprint@dblp#1{\href {http://dblp.uni-trier.de/rec/bibtex/#1.xml}
  {dblp:#1}}
\def\mn@eprint@#1:#2:#3:#4\@nil{\def\@tempa {#1}\def\@tempb {#2}\def\@tempc
  {#3}\ifx \@tempc \@empty \let \@tempc \@tempb \let \@tempb \@tempa \fi \ifx
  \@tempb \@empty \def\@tempb {arXiv}\fi \@ifundefined
  {mn@eprint@\@tempb}{\@tempb:\@tempc}{\expandafter \expandafter \csname
  mn@eprint@\@tempb\endcsname \expandafter{\@tempc}}}

\bibitem[\protect\citeauthoryear{{Allam}, {Lin}, {Tucker}, {Buckley-Geer},
  {Kubik}, {Diehl}, {Annis}  \& {Frieman}}{{Allam} et~al.}{2007}]{Allam07}
{Allam} S.~S.,  {Lin} H.,  {Tucker} D.,  {Buckley-Geer} E.,  {Kubik} D.,
  {Diehl} T.,  {Annis} J.,   {Frieman} J.,  2007, in American Astronomical
  Society Meeting Abstracts. p. 99.02

\bibitem[\protect\citeauthoryear{{Asano}, {Takeuchi}, {Hirashita}  \&
  {Inoue}}{{Asano} et~al.}{2013}]{Asano13}
{Asano} R.~S.,  {Takeuchi} T.~T.,  {Hirashita} H.,   {Inoue} A.~K.,  2013,
  \mn@doi [Earth, Planets, and Space] {10.5047/eps.2012.04.014}, \href
  {https://ui.adsabs.harvard.edu/abs/2013EP&S...65..213A} {65, 213}

\bibitem[\protect\citeauthoryear{{Asplund}, {Grevesse}, {Sauval}  \&
  {Scott}}{{Asplund} et~al.}{2009}]{Asplund09}
{Asplund} M.,  {Grevesse} N.,  {Sauval} A.~J.,   {Scott} P.,  2009, \mn@doi
  [\araa] {10.1146/annurev.astro.46.060407.145222}, \href
  {https://ui.adsabs.harvard.edu/abs/2009ARA&A..47..481A} {47, 481}

\bibitem[\protect\citeauthoryear{{Baker}, {Tacconi}, {Genzel}, {Lehnert}  \&
  {Lutz}}{{Baker} et~al.}{2004}]{Baker04}
{Baker} A.~J.,  {Tacconi} L.~J.,  {Genzel} R.,  {Lehnert} M.~D.,   {Lutz} D.,
  2004, \mn@doi [\apj] {10.1086/381798}, \href
  {https://ui.adsabs.harvard.edu/abs/2004ApJ...604..125B} {604, 125}

\bibitem[\protect\citeauthoryear{{Bastian}}{{Bastian}}{2008}]{Bastian08}
{Bastian} N.,  2008, \mn@doi [\mnras] {10.1111/j.1365-2966.2008.13775.x}, \href
  {http://adsabs.harvard.edu/abs/2008MNRAS.390..759B} {390, 759}

\bibitem[\protect\citeauthoryear{{Baugh}, {Lacey}, {Frenk}, {Granato}, {Silva},
  {Bressan}, {Benson}  \& {Cole}}{{Baugh} et~al.}{2005}]{Baugh05}
{Baugh} C.~M.,  {Lacey} C.~G.,  {Frenk} C.~S.,  {Granato} G.~L.,  {Silva} L.,
  {Bressan} A.,  {Benson} A.~J.,   {Cole} S.,  2005, \mn@doi [\mnras]
  {10.1111/j.1365-2966.2004.08553.x}, \href
  {http://adsabs.harvard.edu/abs/2005MNRAS.356.1191B} {356, 1191}

\bibitem[\protect\citeauthoryear{{Bayliss}, {Rigby}, {Sharon}, {Wuyts},
  {Florian}, {Gladders}, {Johnson}  \& {Oguri}}{{Bayliss}
  et~al.}{2014}]{Bayliss14}
{Bayliss} M.~B.,  {Rigby} J.~R.,  {Sharon} K.,  {Wuyts} E.,  {Florian} M.,
  {Gladders} M.~D.,  {Johnson} T.,   {Oguri} M.,  2014, \mn@doi [\apj]
  {10.1088/0004-637X/790/2/144}, \href
  {http://adsabs.harvard.edu/abs/2014ApJ...790..144B} {790, 144}

\bibitem[\protect\citeauthoryear{{Belokurov} et~al.,}{{Belokurov}
  et~al.}{2007}]{Belokurov07}
{Belokurov} V.,  et~al., 2007, \mn@doi [\apjl] {10.1086/524948}, \href
  {https://ui.adsabs.harvard.edu/abs/2007ApJ...671L...9B} {671, L9}

\bibitem[\protect\citeauthoryear{{Bianchi} \& {Schneider}}{{Bianchi} \&
  {Schneider}}{2007}]{Bianchi07}
{Bianchi} S.,  {Schneider} R.,  2007, \mn@doi [\mnras]
  {10.1111/j.1365-2966.2007.11829.x}, \href
  {https://ui.adsabs.harvard.edu/abs/2007MNRAS.378..973B} {378, 973}

\bibitem[\protect\citeauthoryear{{Bradamante}, {Matteucci}  \&
  {D'Ercole}}{{Bradamante} et~al.}{1998}]{Bradamante98}
{Bradamante} F.,  {Matteucci} F.,   {D'Ercole} A.,  1998, \aap, \href
  {https://ui.adsabs.harvard.edu/abs/1998A&A...337..338B} {337, 338}

\bibitem[\protect\citeauthoryear{{Calura} \& {Matteucci}}{{Calura} \&
  {Matteucci}}{2006}]{Calura06}
{Calura} F.,  {Matteucci} F.,  2006, \mn@doi [\apj] {10.1086/508147}, \href
  {http://adsabs.harvard.edu/abs/2006ApJ...652..889C} {652, 889}

\bibitem[\protect\citeauthoryear{{Calura} \& {Menci}}{{Calura} \&
  {Menci}}{2009}]{Calura09}
{Calura} F.,  {Menci} N.,  2009, \mn@doi [\mnras]
  {10.1111/j.1365-2966.2009.15440.x}, \href
  {http://adsabs.harvard.edu/abs/2009MNRAS.400.1347C} {400, 1347}

\bibitem[\protect\citeauthoryear{{Calura}, {Pipino}  \& {Matteucci}}{{Calura}
  et~al.}{2008}]{Calura08}
{Calura} F.,  {Pipino} A.,   {Matteucci} F.,  2008, \mn@doi [\aap]
  {10.1051/0004-6361:20078090}, \href
  {https://ui.adsabs.harvard.edu/abs/2008A&A...479..669C} {479, 669}

\bibitem[\protect\citeauthoryear{{Calura}, {Recchi}, {Matteucci}  \&
  {Kroupa}}{{Calura} et~al.}{2010}]{Calura10}
{Calura} F.,  {Recchi} S.,  {Matteucci} F.,   {Kroupa} P.,  2010, \mn@doi
  [\mnras] {10.1111/j.1365-2966.2010.16803.x}, \href
  {http://adsabs.harvard.edu/abs/2010MNRAS.406.1985C} {406, 1985}

\bibitem[\protect\citeauthoryear{{Calura}, {Gilli}, {Vignali}, {Pozzi},
  {Pipino}  \& {Matteucci}}{{Calura} et~al.}{2014}]{Calura14}
{Calura} F.,  {Gilli} R.,  {Vignali} C.,  {Pozzi} F.,  {Pipino} A.,
  {Matteucci} F.,  2014, \mn@doi [\mnras] {10.1093/mnras/stt2329}, \href
  {https://ui.adsabs.harvard.edu/abs/2014MNRAS.438.2765C} {438, 2765}

\bibitem[\protect\citeauthoryear{{Calura} et~al.,}{{Calura}
  et~al.}{2017}]{Calura17}
{Calura} F.,  et~al., 2017, \mn@doi [\mnras] {10.1093/mnras/stw2749}, \href
  {https://ui.adsabs.harvard.edu/abs/2017MNRAS.465...54C} {465, 54}

\bibitem[\protect\citeauthoryear{{Cenarro}, {Gorgas}, {Vazdekis}, {Cardiel}  \&
  {Peletier}}{{Cenarro} et~al.}{2003}]{Cenarro03}
{Cenarro} A.~J.,  {Gorgas} J.,  {Vazdekis} A.,  {Cardiel} N.,   {Peletier}
  R.~F.,  2003, \mn@doi [\mnras] {10.1046/j.1365-8711.2003.06360.x}, \href
  {https://ui.adsabs.harvard.edu/abs/2003MNRAS.339L..12C} {339, L12}

\bibitem[\protect\citeauthoryear{{Chiappini}, {Romano}  \&
  {Matteucci}}{{Chiappini} et~al.}{2003}]{Chiappini03}
{Chiappini} C.,  {Romano} D.,   {Matteucci} F.,  2003, \mn@doi [\mnras]
  {10.1046/j.1365-8711.2003.06154.x}, \href
  {https://ui.adsabs.harvard.edu/abs/2003MNRAS.339...63C} {339, 63}

\bibitem[\protect\citeauthoryear{{Christensen} et~al.,}{{Christensen}
  et~al.}{2012a}]{Christensen12a}
{Christensen} L.,  et~al., 2012a, \mn@doi [\mnras]
  {10.1111/j.1365-2966.2012.22006.x}, \href
  {http://adsabs.harvard.edu/abs/2012MNRAS.427.1953C} {427, 1953}

\bibitem[\protect\citeauthoryear{{Christensen} et~al.,}{{Christensen}
  et~al.}{2012b}]{Christensen12b}
{Christensen} L.,  et~al., 2012b, \mn@doi [\mnras]
  {10.1111/j.1365-2966.2012.22007.x}, \href
  {http://adsabs.harvard.edu/abs/2012MNRAS.427.1973C} {427, 1973}

\bibitem[\protect\citeauthoryear{{Conroy} \& {van Dokkum}}{{Conroy} \& {van
  Dokkum}}{2012a}]{Conroy12_2}
{Conroy} C.,  {van Dokkum} P.,  2012a, \mn@doi [\apj]
  {10.1088/0004-637X/747/1/69}, \href
  {https://ui.adsabs.harvard.edu/abs/2012ApJ...747...69C} {747, 69}

\bibitem[\protect\citeauthoryear{{Conroy} \& {van Dokkum}}{{Conroy} \& {van
  Dokkum}}{2012b}]{Conroy12}
{Conroy} C.,  {van Dokkum} P.~G.,  2012b, \mn@doi [\apj]
  {10.1088/0004-637X/760/1/71}, \href
  {http://adsabs.harvard.edu/abs/2012ApJ...760...71C} {760, 71}

\bibitem[\protect\citeauthoryear{{Dav{\'e}}}{{Dav{\'e}}}{2008}]{Dave08}
{Dav{\'e}} R.,  2008, \mn@doi [\mnras] {10.1111/j.1365-2966.2008.12866.x},
  \href {http://adsabs.harvard.edu/abs/2008MNRAS.385..147D} {385, 147}

\bibitem[\protect\citeauthoryear{{De Masi}, {Matteucci}  \& {Vincenzo}}{{De
  Masi} et~al.}{2018}]{DeMasi18}
{De Masi} C.,  {Matteucci} F.,   {Vincenzo} F.,  2018, \mn@doi [\mnras]
  {10.1093/mnras/stx3044}, \href
  {http://adsabs.harvard.edu/abs/2018MNRAS.474.5259D} {474, 5259}

\bibitem[\protect\citeauthoryear{{Dell'Agli}, {Garc{\'\i}a-Hern{\'a}ndez},
  {Schneider}, {Ventura}, {La Franca}, {Valiante}, {Marini}  \& {Di
  Criscienzo}}{{Dell'Agli} et~al.}{2017}]{DellAgli17}
{Dell'Agli} F.,  {Garc{\'\i}a-Hern{\'a}ndez} D.~A.,  {Schneider} R.,  {Ventura}
  P.,  {La Franca} F.,  {Valiante} R.,  {Marini} E.,   {Di Criscienzo} M.,
  2017, \mn@doi [\mnras] {10.1093/mnras/stx387}, \href
  {https://ui.adsabs.harvard.edu/abs/2017MNRAS.467.4431D} {467, 4431}

\bibitem[\protect\citeauthoryear{{Dessauges-Zavadsky}, {D'Odorico}, {Schaerer},
  {Modigliani}, {Tapken}  \& {Vernet}}{{Dessauges-Zavadsky}
  et~al.}{2010}]{Dessauges10}
{Dessauges-Zavadsky} M.,  {D'Odorico} S.,  {Schaerer} D.,  {Modigliani} A.,
  {Tapken} C.,   {Vernet} J.,  2010, \mn@doi [\aap]
  {10.1051/0004-6361/200913337}, \href
  {http://adsabs.harvard.edu/abs/2010A%26A...510A..26D} {510, A26}

\bibitem[\protect\citeauthoryear{{Dwek}}{{Dwek}}{1998}]{Dwek98}
{Dwek} E.,  1998, \mn@doi [\apj] {10.1086/305829}, \href
  {https://ui.adsabs.harvard.edu/abs/1998ApJ...501..643D} {501, 643}

\bibitem[\protect\citeauthoryear{{Dye}, {Evans}, {Belokurov}, {Warren}  \&
  {Hewett}}{{Dye} et~al.}{2008}]{Dye08}
{Dye} S.,  {Evans} N.~W.,  {Belokurov} V.,  {Warren} S.~J.,   {Hewett} P.,
  2008, \mn@doi [\mnras] {10.1111/j.1365-2966.2008.13401.x}, \href
  {https://ui.adsabs.harvard.edu/abs/2008MNRAS.388..384D} {388, 384}

\bibitem[\protect\citeauthoryear{{Ferrara}, {Viti}  \& {Ceccarelli}}{{Ferrara}
  et~al.}{2016}]{Ferrara16}
{Ferrara} A.,  {Viti} S.,   {Ceccarelli} C.,  2016, \mn@doi [\mnras]
  {10.1093/mnrasl/slw165}, \href
  {https://ui.adsabs.harvard.edu/abs/2016MNRAS.463L.112F} {463, L112}

\bibitem[\protect\citeauthoryear{{Ferreras}, {Weidner}, {Vazdekis}  \& {La
  Barbera}}{{Ferreras} et~al.}{2015}]{Ferreras15}
{Ferreras} I.,  {Weidner} C.,  {Vazdekis} A.,   {La Barbera} F.,  2015, \mn@doi
  [\mnras] {10.1093/mnrasl/slv003}, \href
  {http://adsabs.harvard.edu/abs/2015MNRAS.448L..82F} {448, L82}

\bibitem[\protect\citeauthoryear{{Finkelstein}, {Papovich}, {Rudnick}, {Egami},
  {Le Floc'h}, {Rieke}, {Rigby}  \& {Willmer}}{{Finkelstein}
  et~al.}{2009}]{Finkelstein09}
{Finkelstein} S.~L.,  {Papovich} C.,  {Rudnick} G.,  {Egami} E.,  {Le Floc'h}
  E.,  {Rieke} M.~J.,  {Rigby} J.~R.,   {Willmer} C.~N.~A.,  2009, \mn@doi
  [\apj] {10.1088/0004-637X/700/1/376}, \href
  {http://adsabs.harvard.edu/abs/2009ApJ...700..376F} {700, 376}

\bibitem[\protect\citeauthoryear{{Fran{\c c}ois}, {Matteucci}, {Cayrel},
  {Spite}, {Spite}  \& {Chiappini}}{{Fran{\c c}ois} et~al.}{2004}]{Francois04}
{Fran{\c c}ois} P.,  {Matteucci} F.,  {Cayrel} R.,  {Spite} M.,  {Spite} F.,
  {Chiappini} C.,  2004, \mn@doi [\aap] {10.1051/0004-6361:20034140}, \href
  {http://adsabs.harvard.edu/abs/2004A%26A...421..613F} {421, 613}

\bibitem[\protect\citeauthoryear{{Gall} \& {Hjorth}}{{Gall} \&
  {Hjorth}}{2018}]{Gall18}
{Gall} C.,  {Hjorth} J.,  2018, \mn@doi [\apj] {10.3847/1538-4357/aae520},
  \href {https://ui.adsabs.harvard.edu/abs/2018ApJ...868...62G} {868, 62}

\bibitem[\protect\citeauthoryear{{Gall}, {Hjorth}  \& {Andersen}}{{Gall}
  et~al.}{2011}]{Gall11}
{Gall} C.,  {Hjorth} J.,   {Andersen} A.~C.,  2011, \mn@doi [\aapr]
  {10.1007/s00159-011-0043-7}, \href
  {https://ui.adsabs.harvard.edu/abs/2011A&ARv..19...43G} {19, 43}

\bibitem[\protect\citeauthoryear{{Garnett}, {Skillman}, {Dufour}, {Peimbert},
  {Torres-Peimbert}, {Terlevich}, {Terlevich}  \& {Shields}}{{Garnett}
  et~al.}{1995}]{Garnett95}
{Garnett} D.~R.,  {Skillman} E.~D.,  {Dufour} R.~J.,  {Peimbert} M.,
  {Torres-Peimbert} S.,  {Terlevich} R.,  {Terlevich} E.,   {Shields} G.~A.,
  1995, \mn@doi [\apj] {10.1086/175503}, \href
  {https://ui.adsabs.harvard.edu/abs/1995ApJ...443...64G} {443, 64}

\bibitem[\protect\citeauthoryear{{Gibson} \& {Matteucci}}{{Gibson} \&
  {Matteucci}}{1997}]{Gibson97}
{Gibson} B.~K.,  {Matteucci} F.,  1997, \mn@doi [\mnras]
  {10.1093/mnras/291.1.L8}, \href
  {http://adsabs.harvard.edu/abs/1997MNRAS.291L...8G} {291, L8}

\bibitem[\protect\citeauthoryear{{Gioannini}, {Matteucci}, {Vladilo}  \&
  {Calura}}{{Gioannini} et~al.}{2017}]{Gioannini17}
{Gioannini} L.,  {Matteucci} F.,  {Vladilo} G.,   {Calura} F.,  2017, \mn@doi
  [\mnras] {10.1093/mnras/stw2343}, \href
  {http://adsabs.harvard.edu/abs/2017MNRAS.464..985G} {464, 985}

\bibitem[\protect\citeauthoryear{{Gomez} et~al.,}{{Gomez}
  et~al.}{2012}]{Gomez12}
{Gomez} H.~L.,  et~al., 2012, \mn@doi [\apj] {10.1088/0004-637X/760/1/96},
  \href {https://ui.adsabs.harvard.edu/abs/2012ApJ...760...96G} {760, 96}

\bibitem[\protect\citeauthoryear{{Grieco}, {Matteucci}, {Calura}, {Boissier},
  {Longo}  \& {D'Elia}}{{Grieco} et~al.}{2014}]{Grieco14}
{Grieco} V.,  {Matteucci} F.,  {Calura} F.,  {Boissier} S.,  {Longo} F.,
  {D'Elia} V.,  2014, \mn@doi [\mnras] {10.1093/mnras/stu1500}, \href
  {https://ui.adsabs.harvard.edu/abs/2014MNRAS.444.1054G} {444, 1054}

\bibitem[\protect\citeauthoryear{{Hainline}, {Shapley}, {Kornei}, {Pettini},
  {Buckley-Geer}, {Allam}  \& {Tucker}}{{Hainline} et~al.}{2009}]{Hainline09}
{Hainline} K.~N.,  {Shapley} A.~E.,  {Kornei} K.~A.,  {Pettini} M.,
  {Buckley-Geer} E.,  {Allam} S.~S.,   {Tucker} D.~L.,  2009, \mn@doi [\apj]
  {10.1088/0004-637X/701/1/52}, \href
  {https://ui.adsabs.harvard.edu/abs/2009ApJ...701...52H} {701, 52}

\bibitem[\protect\citeauthoryear{{Israelian}, {Ecuvillon}, {Rebolo},
  {Garc{\'\i}a-L{\'o}pez}, {Bonifacio}  \& {Molaro}}{{Israelian}
  et~al.}{2004}]{Israelian04}
{Israelian} G.,  {Ecuvillon} A.,  {Rebolo} R.,  {Garc{\'\i}a-L{\'o}pez} R.,
  {Bonifacio} P.,   {Molaro} P.,  2004, \mn@doi [\aap]
  {10.1051/0004-6361:20047132}, \href
  {https://ui.adsabs.harvard.edu/abs/2004A&A...421..649I} {421, 649}

\bibitem[\protect\citeauthoryear{{Iwamoto}, {Brachwitz}, {Nomoto}, {Kishimoto},
  {Umeda}, {Hix}  \& {Thielemann}}{{Iwamoto} et~al.}{1999}]{Iwamoto99}
{Iwamoto} K.,  {Brachwitz} F.,  {Nomoto} K.,  {Kishimoto} N.,  {Umeda} H.,
  {Hix} W.~R.,   {Thielemann} F.-K.,  1999, \mn@doi [\apjs] {10.1086/313278},
  \href {https://ui.adsabs.harvard.edu/abs/1999ApJS..125..439I} {125, 439}

\bibitem[\protect\citeauthoryear{{Je{\v{r}}{\'a}bkov{\'a}}, {Kroupa},
  {Dabringhausen}, {Hilker}  \& {Bekki}}{{Je{\v{r}}{\'a}bkov{\'a}}
  et~al.}{2017}]{Jerabkova17}
{Je{\v{r}}{\'a}bkov{\'a}} T.,  {Kroupa} P.,  {Dabringhausen} J.,  {Hilker} M.,
   {Bekki} K.,  2017, \mn@doi [\aap] {10.1051/0004-6361/201731240}, \href
  {https://ui.adsabs.harvard.edu/abs/2017A&A...608A..53J} {608, A53}

\bibitem[\protect\citeauthoryear{{Je{\v{r}}{\'a}bkov{\'a}}, {Hasani Zonoozi},
  {Kroupa}, {Beccari}, {Yan}, {Vazdekis}  \& {Zhang}}{{Je{\v{r}}{\'a}bkov{\'a}}
  et~al.}{2018}]{Jerabkova18}
{Je{\v{r}}{\'a}bkov{\'a}} T.,  {Hasani Zonoozi} A.,  {Kroupa} P.,  {Beccari}
  G.,  {Yan} Z.,  {Vazdekis} A.,   {Zhang} Z.~Y.,  2018, \mn@doi [\aap]
  {10.1051/0004-6361/201833055}, \href
  {https://ui.adsabs.harvard.edu/abs/2018A&A...620A..39J} {620, A39}

\bibitem[\protect\citeauthoryear{{Kennicutt}}{{Kennicutt}}{1998}]{Kennicutt98}
{Kennicutt} Jr. R.~C.,  1998, \mn@doi [\apj] {10.1086/305588}, \href
  {http://adsabs.harvard.edu/abs/1998ApJ...498..541K} {498, 541}

\bibitem[\protect\citeauthoryear{{Kroupa}}{{Kroupa}}{2001}]{Kroupa01}
{Kroupa} P.,  2001, \mn@doi [\mnras] {10.1046/j.1365-8711.2001.04022.x}, \href
  {http://adsabs.harvard.edu/abs/2001MNRAS.322..231K} {322, 231}

\bibitem[\protect\citeauthoryear{{Kroupa}}{{Kroupa}}{2002}]{Kroupa02}
{Kroupa} P.,  2002, \mn@doi [Science] {10.1126/science.1067524}, \href
  {http://adsabs.harvard.edu/abs/2002Sci...295...82K} {295, 82}

\bibitem[\protect\citeauthoryear{{Kroupa} \& {Weidner}}{{Kroupa} \&
  {Weidner}}{2003}]{Kroupa03}
{Kroupa} P.,  {Weidner} C.,  2003, \mn@doi [\apj] {10.1086/379105}, \href
  {http://adsabs.harvard.edu/abs/2003ApJ...598.1076K} {598, 1076}

\bibitem[\protect\citeauthoryear{{La Barbera}, {Ferreras}, {Vazdekis}, {de la
  Rosa}, {de Carvalho}, {Trevisan}, {Falc{\'o}n-Barroso}  \&
  {Ricciardelli}}{{La Barbera} et~al.}{2013}]{LaBarbera13}
{La Barbera} F.,  {Ferreras} I.,  {Vazdekis} A.,  {de la Rosa} I.~G.,  {de
  Carvalho} R.~R.,  {Trevisan} M.,  {Falc{\'o}n-Barroso} J.,   {Ricciardelli}
  E.,  2013, \mn@doi [\mnras] {10.1093/mnras/stt943}, \href
  {https://ui.adsabs.harvard.edu/abs/2013MNRAS.433.3017L} {433, 3017}

\bibitem[\protect\citeauthoryear{{Lacchin}, {Matteucci}, {Vincenzo}  \&
  {Palla}}{{Lacchin} et~al.}{2019}]{Lacchin20}
{Lacchin} E.,  {Matteucci} F.,  {Vincenzo} F.,   {Palla} M.,  2019, arXiv
  e-prints, \href {https://ui.adsabs.harvard.edu/abs/2019arXiv191108450L} {p.
  arXiv:1911.08450}

\bibitem[\protect\citeauthoryear{{Lada} \& {Lada}}{{Lada} \&
  {Lada}}{2003}]{Lada03}
{Lada} C.~J.,  {Lada} E.~A.,  2003, \mn@doi [\araa]
  {10.1146/annurev.astro.41.011802.094844}, \href
  {http://adsabs.harvard.edu/abs/2003ARA%26A..41...57L} {41, 57}

\bibitem[\protect\citeauthoryear{{Larson}}{{Larson}}{1998}]{Larson98}
{Larson} R.~B.,  1998, \mn@doi [\mnras] {10.1046/j.1365-8711.1998.02045.x},
  \href {http://adsabs.harvard.edu/abs/1998MNRAS.301..569L} {301, 569}

\bibitem[\protect\citeauthoryear{{L{\"u}ghausen}, {Famaey}, {Kroupa}, {Angus},
  {Combes}, {Gentile}, {Tiret}  \& {Zhao}}{{L{\"u}ghausen}
  et~al.}{2013}]{Lueghausen13}
{L{\"u}ghausen} F.,  {Famaey} B.,  {Kroupa} P.,  {Angus} G.,  {Combes} F.,
  {Gentile} G.,  {Tiret} O.,   {Zhao} H.,  2013, \mn@doi [\mnras]
  {10.1093/mnras/stt639}, \href
  {https://ui.adsabs.harvard.edu/abs/2013MNRAS.432.2846L} {432, 2846}

\bibitem[\protect\citeauthoryear{{L{\"u}ghausen}, {Famaey}  \&
  {Kroupa}}{{L{\"u}ghausen} et~al.}{2015}]{Lueghausen15}
{L{\"u}ghausen} F.,  {Famaey} B.,   {Kroupa} P.,  2015, \mn@doi [Canadian
  Journal of Physics] {10.1139/cjp-2014-0168}, \href
  {https://ui.adsabs.harvard.edu/abs/2015CaJPh..93..232L} {93, 232}

\bibitem[\protect\citeauthoryear{{Massey} \& {Hunter}}{{Massey} \&
  {Hunter}}{1998}]{Massey98}
{Massey} P.,  {Hunter} D.~A.,  1998, \mn@doi [\apj] {10.1086/305126}, \href
  {http://adsabs.harvard.edu/abs/1998ApJ...493..180M} {493, 180}

\bibitem[\protect\citeauthoryear{{Matteucci}}{{Matteucci}}{1986}]{Matteucci86}
{Matteucci} F.,  1986, \mn@doi [\mnras] {10.1093/mnras/221.4.911}, \href
  {http://adsabs.harvard.edu/abs/1986MNRAS.221..911M} {221, 911}

\bibitem[\protect\citeauthoryear{{Matteucci}}{{Matteucci}}{1994}]{Matteucci94}
{Matteucci} F.,  1994, \aap, \href
  {http://adsabs.harvard.edu/abs/1994A%26A...288...57M} {288, 57}

\bibitem[\protect\citeauthoryear{{Matteucci}}{{Matteucci}}{2012}]{Matteucci12}
{Matteucci} F.,  2012, {Chemical Evolution of Galaxies},
  \mn@doi{10.1007/978-3-642-22491-1.
}

\bibitem[\protect\citeauthoryear{{Matteucci} \& {Greggio}}{{Matteucci} \&
  {Greggio}}{1986}]{MatteucciGreggio86}
{Matteucci} F.,  {Greggio} L.,  1986, \aap, \href
  {http://adsabs.harvard.edu/abs/1986A%26A...154..279M} {154, 279}

\bibitem[\protect\citeauthoryear{{Matteucci} \& {Pipino}}{{Matteucci} \&
  {Pipino}}{2002}]{Matteucci02}
{Matteucci} F.,  {Pipino} A.,  2002, \mn@doi [\apjl] {10.1086/340771}, \href
  {https://ui.adsabs.harvard.edu/abs/2002ApJ...569L..69M} {569, L69}

\bibitem[\protect\citeauthoryear{{Matteucci} \& {Recchi}}{{Matteucci} \&
  {Recchi}}{2001}]{Matteucci01}
{Matteucci} F.,  {Recchi} S.,  2001, \mn@doi [\apj] {10.1086/322472}, \href
  {http://adsabs.harvard.edu/abs/2001ApJ...558..351M} {558, 351}

\bibitem[\protect\citeauthoryear{{Mattsson}}{{Mattsson}}{2011}]{Mattsson11}
{Mattsson} L.,  2011, \mn@doi [\mnras] {10.1111/j.1365-2966.2011.18447.x},
  \href {https://ui.adsabs.harvard.edu/abs/2011MNRAS.414..781M} {414, 781}

\bibitem[\protect\citeauthoryear{{Mattsson}}{{Mattsson}}{2015}]{Matsson15}
{Mattsson} L.,  2015, arXiv e-prints, \href
  {https://ui.adsabs.harvard.edu/abs/2015arXiv150504758M} {p. arXiv:1505.04758}

\bibitem[\protect\citeauthoryear{{Megeath} et~al.,}{{Megeath}
  et~al.}{2016}]{Megheat16}
{Megeath} S.~T.,  et~al., 2016, \mn@doi [\aj] {10.3847/0004-6256/151/1/5},
  \href {https://ui.adsabs.harvard.edu/abs/2016AJ....151....5M} {151, 5}

\bibitem[\protect\citeauthoryear{{Meneghetti} et~al.,}{{Meneghetti}
  et~al.}{2017}]{Meneghetti17}
{Meneghetti} M.,  et~al., 2017, \mn@doi [\mnras] {10.1093/mnras/stx2064}, \href
  {https://ui.adsabs.harvard.edu/abs/2017MNRAS.472.3177M} {472, 3177}

\bibitem[\protect\citeauthoryear{{Meynet} \& {Maeder}}{{Meynet} \&
  {Maeder}}{2002}]{Meynet02}
{Meynet} G.,  {Maeder} A.,  2002, \mn@doi [\aap] {10.1051/0004-6361:20020755},
  \href {https://ui.adsabs.harvard.edu/abs/2002A&A...390..561M} {390, 561}

\bibitem[\protect\citeauthoryear{{Nittler}, {O'D. Alexander}, {Liu}  \&
  {Wang}}{{Nittler} et~al.}{2018}]{Nittler18}
{Nittler} L.~R.,  {O'D. Alexander} C.~M.,  {Liu} N.,   {Wang} J.,  2018,
  \mn@doi [\apjl] {10.3847/2041-8213/aab61f}, \href
  {https://ui.adsabs.harvard.edu/abs/2018ApJ...856L..24N} {856, L24}

\bibitem[\protect\citeauthoryear{{Nozawa}, {Maeda}, {Kozasa}, {Tanaka},
  {Nomoto}  \& {Umeda}}{{Nozawa} et~al.}{2011}]{Nozawa11}
{Nozawa} T.,  {Maeda} K.,  {Kozasa} T.,  {Tanaka} M.,  {Nomoto} K.,   {Umeda}
  H.,  2011, \mn@doi [\apj] {10.1088/0004-637X/736/1/45}, \href
  {https://ui.adsabs.harvard.edu/abs/2011ApJ...736...45N} {736, 45}

\bibitem[\protect\citeauthoryear{{Palla}, {Matteucci}, {Calura}  \&
  {Longo}}{{Palla} et~al.}{2020}]{Palla20}
{Palla} M.,  {Matteucci} F.,  {Calura} F.,   {Longo} F.,  2020, \mn@doi [\apj]
  {10.3847/1538-4357/ab6080}, \href
  {https://ui.adsabs.harvard.edu/abs/2020ApJ...889....4P} {889, 4}

\bibitem[\protect\citeauthoryear{{Pettini}, {Steidel}, {Adelberger},
  {Dickinson}  \& {Giavalisco}}{{Pettini} et~al.}{2000}]{Pettini00}
{Pettini} M.,  {Steidel} C.~C.,  {Adelberger} K.~L.,  {Dickinson} M.,
  {Giavalisco} M.,  2000, \mn@doi [\apj] {10.1086/308176}, \href
  {http://adsabs.harvard.edu/abs/2000ApJ...528...96P} {528, 96}

\bibitem[\protect\citeauthoryear{{Pettini}, {Shapley}, {Steidel}, {Cuby},
  {Dickinson}, {Moorwood}, {Adelberger}  \& {Giavalisco}}{{Pettini}
  et~al.}{2001}]{Pettini01}
{Pettini} M.,  {Shapley} A.~E.,  {Steidel} C.~C.,  {Cuby} J.-G.,  {Dickinson}
  M.,  {Moorwood} A.~F.~M.,  {Adelberger} K.~L.,   {Giavalisco} M.,  2001,
  \mn@doi [\apj] {10.1086/321403}, \href
  {http://adsabs.harvard.edu/abs/2001ApJ...554..981P} {554, 981}

\bibitem[\protect\citeauthoryear{{Pettini}, {Rix}, {Steidel}, {Adelberger},
  {Hunt}  \& {Shapley}}{{Pettini} et~al.}{2002}]{Pettini02}
{Pettini} M.,  {Rix} S.~A.,  {Steidel} C.~C.,  {Adelberger} K.~L.,  {Hunt}
  M.~P.,   {Shapley} A.~E.,  2002, \mn@doi [\apj] {10.1086/339355}, \href
  {http://adsabs.harvard.edu/abs/2002ApJ...569..742P} {569, 742}

\bibitem[\protect\citeauthoryear{{Pflamm-Altenburg}, {Weidner}  \&
  {Kroupa}}{{Pflamm-Altenburg} et~al.}{2007}]{Pflamm07}
{Pflamm-Altenburg} J.,  {Weidner} C.,   {Kroupa} P.,  2007, \mn@doi [\apj]
  {10.1086/523033}, \href {http://adsabs.harvard.edu/abs/2007ApJ...671.1550P}
  {671, 1550}

\bibitem[\protect\citeauthoryear{{Pipino} \& {Matteucci}}{{Pipino} \&
  {Matteucci}}{2004}]{Pipino04}
{Pipino} A.,  {Matteucci} F.,  2004, \mn@doi [\mnras]
  {10.1111/j.1365-2966.2004.07268.x}, \href
  {https://ui.adsabs.harvard.edu/abs/2004MNRAS.347..968P} {347, 968}

\bibitem[\protect\citeauthoryear{{Pipino} \& {Matteucci}}{{Pipino} \&
  {Matteucci}}{2011}]{Pipino11b}
{Pipino} A.,  {Matteucci} F.,  2011, \mn@doi [\aap]
  {10.1051/0004-6361/201016191}, \href
  {https://ui.adsabs.harvard.edu/abs/2011A&A...530A..98P} {530, A98}

\bibitem[\protect\citeauthoryear{{Pipino}, {Matteucci}, {Borgani}  \&
  {Biviano}}{{Pipino} et~al.}{2002}]{Pipino02}
{Pipino} A.,  {Matteucci} F.,  {Borgani} S.,   {Biviano} A.,  2002, \mn@doi
  [\na] {10.1016/S1384-1076(02)00136-7}, \href
  {https://ui.adsabs.harvard.edu/abs/2002NewA....7..227P} {7, 227}

\bibitem[\protect\citeauthoryear{{Pipino}, {Fan}, {Matteucci}, {Calura},
  {Silva}, {Granato}  \& {Maiolino}}{{Pipino} et~al.}{2011}]{Pipino11}
{Pipino} A.,  {Fan} X.~L.,  {Matteucci} F.,  {Calura} F.,  {Silva} L.,
  {Granato} G.,   {Maiolino} R.,  2011, \mn@doi [\aap]
  {10.1051/0004-6361/201014843}, \href
  {http://adsabs.harvard.edu/abs/2011A%26A...525A..61P} {525, A61}

\bibitem[\protect\citeauthoryear{{Pozzi}, {Calura}, {Zamorani}, {Delvecchio},
  {Gruppioni}  \& {Santini}}{{Pozzi} et~al.}{2020}]{Pozzi20}
{Pozzi} F.,  {Calura} F.,  {Zamorani} G.,  {Delvecchio} I.,  {Gruppioni} C.,
  {Santini} P.,  2020, \mn@doi [\mnras] {10.1093/mnras/stz2724}, \href
  {https://ui.adsabs.harvard.edu/abs/2020MNRAS.491.5073P} {491, 5073}

\bibitem[\protect\citeauthoryear{{Quider}, {Pettini}, {Shapley}  \&
  {Steidel}}{{Quider} et~al.}{2009}]{Quider09}
{Quider} A.~M.,  {Pettini} M.,  {Shapley} A.~E.,   {Steidel} C.~C.,  2009,
  \mn@doi [\mnras] {10.1111/j.1365-2966.2009.15234.x}, \href
  {https://ui.adsabs.harvard.edu/abs/2009MNRAS.398.1263Q} {398, 1263}

\bibitem[\protect\citeauthoryear{{Recchi}, {Matteucci}  \& {D'Ercole}}{{Recchi}
  et~al.}{2001}]{Recchi01}
{Recchi} S.,  {Matteucci} F.,   {D'Ercole} A.,  2001, \mn@doi [\mnras]
  {10.1046/j.1365-8711.2001.04189.x}, \href
  {http://adsabs.harvard.edu/abs/2001MNRAS.322..800R} {322, 800}

\bibitem[\protect\citeauthoryear{{Recchi}, {Calura}  \& {Kroupa}}{{Recchi}
  et~al.}{2009}]{Recchi09}
{Recchi} S.,  {Calura} F.,   {Kroupa} P.,  2009, \mn@doi [\aap]
  {10.1051/0004-6361/200811472}, \href
  {http://adsabs.harvard.edu/abs/2009A%26A...499..711R} {499, 711}

\bibitem[\protect\citeauthoryear{{Recchi}, {Calura}, {Gibson}  \&
  {Kroupa}}{{Recchi} et~al.}{2014}]{Recchi14}
{Recchi} S.,  {Calura} F.,  {Gibson} B.~K.,   {Kroupa} P.,  2014, \mn@doi
  [\mnras] {10.1093/mnras/stt1971}, \href
  {http://adsabs.harvard.edu/abs/2014MNRAS.437..994R} {437, 994}

\bibitem[\protect\citeauthoryear{{Rigby}, {Wuyts}, {Gladders}, {Sharon}  \&
  {Becker}}{{Rigby} et~al.}{2011}]{Rigby11}
{Rigby} J.~R.,  {Wuyts} E.,  {Gladders} M.~D.,  {Sharon} K.,   {Becker} G.~D.,
  2011, \mn@doi [\apj] {10.1088/0004-637X/732/1/59}, \href
  {http://adsabs.harvard.edu/abs/2011ApJ...732...59R} {732, 59}

\bibitem[\protect\citeauthoryear{{Romano}, {Karakas}, {Tosi}  \&
  {Matteucci}}{{Romano} et~al.}{2010}]{Romano10}
{Romano} D.,  {Karakas} A.~I.,  {Tosi} M.,   {Matteucci} F.,  2010, \mn@doi
  [\aap] {10.1051/0004-6361/201014483}, \href
  {https://ui.adsabs.harvard.edu/abs/2010A&A...522A..32R} {522, A32}

\bibitem[\protect\citeauthoryear{{Romano}, {Matteucci}, {Zhang}, {Papadopoulos}
   \& {Ivison}}{{Romano} et~al.}{2017}]{Romano17}
{Romano} D.,  {Matteucci} F.,  {Zhang} Z.~Y.,  {Papadopoulos} P.~P.,   {Ivison}
  R.~J.,  2017, \mn@doi [\mnras] {10.1093/mnras/stx1197}, \href
  {https://ui.adsabs.harvard.edu/abs/2017MNRAS.470..401R} {470, 401}

\bibitem[\protect\citeauthoryear{{Rowlands}, {Gomez}, {Dunne},
  {Arag{\'o}n-Salamanca}, {Dye}, {Maddox}, {da Cunha}  \& {van der
  Werf}}{{Rowlands} et~al.}{2014}]{Rowlands14}
{Rowlands} K.,  {Gomez} H.~L.,  {Dunne} L.,  {Arag{\'o}n-Salamanca} A.,  {Dye}
  S.,  {Maddox} S.,  {da Cunha} E.,   {van der Werf} P.,  2014, \mn@doi
  [\mnras] {10.1093/mnras/stu605}, \href
  {https://ui.adsabs.harvard.edu/abs/2014MNRAS.441.1040R} {441, 1040}

\bibitem[\protect\citeauthoryear{{Salpeter}}{{Salpeter}}{1955}]{Salpeter55}
{Salpeter} E.~E.,  1955, \mn@doi [\apj] {10.1086/145971}, \href
  {http://adsabs.harvard.edu/abs/1955ApJ...121..161S} {121, 161}

\bibitem[\protect\citeauthoryear{{Salvador-Rusi{\~n}ol}, {Vazdekis}, {La
  Barbera}, {Beasley}, {Ferreras}, {Negri}  \& {Dalla
  Vecchia}}{{Salvador-Rusi{\~n}ol} et~al.}{2019}]{Salvador19}
{Salvador-Rusi{\~n}ol} N.,  {Vazdekis} A.,  {La Barbera} F.,  {Beasley} M.~A.,
  {Ferreras} I.,  {Negri} A.,   {Dalla Vecchia} C.,  2019, \mn@doi [Nature
  Astronomy] {10.1038/s41550-019-0955-0}, \href
  {https://ui.adsabs.harvard.edu/abs/2019NatAs.tmp....1S} {p.~1}

\bibitem[\protect\citeauthoryear{{Schmidt}}{{Schmidt}}{1959}]{Schmidt59}
{Schmidt} M.,  1959, \mn@doi [\apj] {10.1086/146614}, \href
  {http://adsabs.harvard.edu/abs/1959ApJ...129..243S} {129, 243}

\bibitem[\protect\citeauthoryear{{Seitz}, {Saglia}, {Bender}, {Hopp}, {Belloni}
   \& {Ziegler}}{{Seitz} et~al.}{1998}]{Seitz98}
{Seitz} S.,  {Saglia} R.~P.,  {Bender} R.,  {Hopp} U.,  {Belloni} P.,
  {Ziegler} B.,  1998, \mn@doi [\mnras] {10.1046/j.1365-8711.1998.01443.x},
  \href {https://ui.adsabs.harvard.edu/abs/1998MNRAS.298..945S} {298, 945}

\bibitem[\protect\citeauthoryear{{Shapley}, {Steidel}, {Pettini}  \&
  {Adelberger}}{{Shapley} et~al.}{2003}]{Shapley03}
{Shapley} A.~E.,  {Steidel} C.~C.,  {Pettini} M.,   {Adelberger} K.~L.,  2003,
  \mn@doi [\apj] {10.1086/373922}, \href
  {http://adsabs.harvard.edu/abs/2003ApJ...588...65S} {588, 65}

\bibitem[\protect\citeauthoryear{{Siana}, {Teplitz}, {Chary}, {Colbert}  \&
  {Frayer}}{{Siana} et~al.}{2008}]{Siana08}
{Siana} B.,  {Teplitz} H.~I.,  {Chary} R.-R.,  {Colbert} J.,   {Frayer} D.~T.,
  2008, \mn@doi [\apj] {10.1086/592682}, \href
  {http://adsabs.harvard.edu/abs/2008ApJ...689...59S} {689, 59}

\bibitem[\protect\citeauthoryear{{Spite} et~al.,}{{Spite}
  et~al.}{2005}]{Spite05}
{Spite} M.,  et~al., 2005, \mn@doi [\aap] {10.1051/0004-6361:20041274}, \href
  {https://ui.adsabs.harvard.edu/abs/2005A&A...430..655S} {430, 655}

\bibitem[\protect\citeauthoryear{{Spolaor}, {Kobayashi}, {Forbes}, {Couch}  \&
  {Hau}}{{Spolaor} et~al.}{2010}]{spolaor10}
{Spolaor} M.,  {Kobayashi} C.,  {Forbes} D.~A.,  {Couch} W.~J.,   {Hau} G.
  K.~T.,  2010, \mn@doi [\mnras] {10.1111/j.1365-2966.2010.17080.x}, \href
  {https://ui.adsabs.harvard.edu/abs/2010MNRAS.408..272S} {408, 272}

\bibitem[\protect\citeauthoryear{{Steidel} et~al.,}{{Steidel}
  et~al.}{2014}]{Steidel14}
{Steidel} C.~C.,  et~al., 2014, \mn@doi [\apj] {10.1088/0004-637X/795/2/165},
  \href {https://ui.adsabs.harvard.edu/abs/2014ApJ...795..165S} {795, 165}

\bibitem[\protect\citeauthoryear{{Steidel}, {Strom}, {Pettini}, {Rudie},
  {Reddy}  \& {Trainor}}{{Steidel} et~al.}{2016}]{Steidel16}
{Steidel} C.~C.,  {Strom} A.~L.,  {Pettini} M.,  {Rudie} G.~C.,  {Reddy} N.~A.,
    {Trainor} R.~F.,  2016, \mn@doi [\apj] {10.3847/0004-637X/826/2/159}, \href
  {http://adsabs.harvard.edu/abs/2016ApJ...826..159S} {826, 159}

\bibitem[\protect\citeauthoryear{{Teplitz} et~al.,}{{Teplitz}
  et~al.}{2000}]{Teplitz00}
{Teplitz} H.~I.,  et~al., 2000, \mn@doi [\apjl] {10.1086/312595}, \href
  {http://adsabs.harvard.edu/abs/2000ApJ...533L..65T} {533, L65}

\bibitem[\protect\citeauthoryear{{Thies}, {Pflamm-Altenburg}, {Kroupa}  \&
  {Marks}}{{Thies} et~al.}{2015}]{Thies15}
{Thies} I.,  {Pflamm-Altenburg} J.,  {Kroupa} P.,   {Marks} M.,  2015, \mn@doi
  [\apj] {10.1088/0004-637X/800/1/72}, \href
  {https://ui.adsabs.harvard.edu/abs/2015ApJ...800...72T} {800, 72}

\bibitem[\protect\citeauthoryear{{Thomas}, {Maraston}, {Schawinski}, {Sarzi}
  \& {Silk}}{{Thomas} et~al.}{2010}]{Thomas10}
{Thomas} D.,  {Maraston} C.,  {Schawinski} K.,  {Sarzi} M.,   {Silk} J.,  2010,
  \mn@doi [\mnras] {10.1111/j.1365-2966.2010.16427.x}, \href
  {https://ui.adsabs.harvard.edu/abs/2010MNRAS.404.1775T} {404, 1775}

\bibitem[\protect\citeauthoryear{{Valiante}, {Schneider}, {Salvadori}  \&
  {Gallerani}}{{Valiante} et~al.}{2014}]{Valiante14}
{Valiante} R.,  {Schneider} R.,  {Salvadori} S.,   {Gallerani} S.,  2014,
  \mn@doi [\mnras] {10.1093/mnras/stu1613}, \href
  {https://ui.adsabs.harvard.edu/abs/2014MNRAS.444.2442V} {444, 2442}

\bibitem[\protect\citeauthoryear{{Vanzella} et~al.,}{{Vanzella}
  et~al.}{2017}]{Vanzella17}
{Vanzella} E.,  et~al., 2017, \mn@doi [\mnras] {10.1093/mnras/stx351}, \href
  {https://ui.adsabs.harvard.edu/abs/2017MNRAS.467.4304V} {467, 4304}

\bibitem[\protect\citeauthoryear{{Vincenzo} \& {Kobayashi}}{{Vincenzo} \&
  {Kobayashi}}{2018}]{Vincenzo18}
{Vincenzo} F.,  {Kobayashi} C.,  2018, \mn@doi [\aap]
  {10.1051/0004-6361/201732395}, \href
  {https://ui.adsabs.harvard.edu/abs/2018A&A...610L..16V} {610, L16}

\bibitem[\protect\citeauthoryear{{Vincenzo}, {Matteucci}, {Recchi}, {Calura},
  {McWilliam}  \& {Lanfranchi}}{{Vincenzo} et~al.}{2015}]{Vincenzo15}
{Vincenzo} F.,  {Matteucci} F.,  {Recchi} S.,  {Calura} F.,  {McWilliam} A.,
  {Lanfranchi} G.~A.,  2015, \mn@doi [\mnras] {10.1093/mnras/stv357}, \href
  {http://adsabs.harvard.edu/abs/2015MNRAS.449.1327V} {449, 1327}

\bibitem[\protect\citeauthoryear{{Vincenzo}, {Belfiore}, {Maiolino},
  {Matteucci}  \& {Ventura}}{{Vincenzo} et~al.}{2016}]{Vincenzo16}
{Vincenzo} F.,  {Belfiore} F.,  {Maiolino} R.,  {Matteucci} F.,   {Ventura} P.,
   2016, \mn@doi [\mnras] {10.1093/mnras/stw532}, \href
  {https://ui.adsabs.harvard.edu/abs/2016MNRAS.458.3466V} {458, 3466}

\bibitem[\protect\citeauthoryear{{Vladilo}, {Gioannini}, {Matteucci}  \&
  {Palla}}{{Vladilo} et~al.}{2018}]{Vladilo18}
{Vladilo} G.,  {Gioannini} L.,  {Matteucci} F.,   {Palla} M.,  2018, \mn@doi
  [\apj] {10.3847/1538-4357/aae8dc}, \href
  {http://adsabs.harvard.edu/abs/2018ApJ...868..127V} {868, 127}

\bibitem[\protect\citeauthoryear{{Weidner} \& {Kroupa}}{{Weidner} \&
  {Kroupa}}{2004}]{Weidner04}
{Weidner} C.,  {Kroupa} P.,  2004, \mn@doi [\mnras]
  {10.1111/j.1365-2966.2004.07340.x}, \href
  {http://adsabs.harvard.edu/abs/2004MNRAS.348..187W} {348, 187}

\bibitem[\protect\citeauthoryear{{Weidner} \& {Kroupa}}{{Weidner} \&
  {Kroupa}}{2005}]{Weidner05}
{Weidner} C.,  {Kroupa} P.,  2005, \mn@doi [\apj] {10.1086/429867}, \href
  {http://adsabs.harvard.edu/abs/2005ApJ...625..754W} {625, 754}

\bibitem[\protect\citeauthoryear{{Weidner}, {Kroupa}  \& {Larsen}}{{Weidner}
  et~al.}{2004}]{Weidner04_}
{Weidner} C.,  {Kroupa} P.,   {Larsen} S.~S.,  2004, \mn@doi [\mnras]
  {10.1111/j.1365-2966.2004.07758.x}, \href
  {https://ui.adsabs.harvard.edu/abs/2004MNRAS.350.1503W} {350, 1503}

\bibitem[\protect\citeauthoryear{{Weidner}, {Kroupa}  \&
  {Pflamm-Altenburg}}{{Weidner} et~al.}{2011}]{Weidner11}
{Weidner} C.,  {Kroupa} P.,   {Pflamm-Altenburg} J.,  2011, \mn@doi [\mnras]
  {10.1111/j.1365-2966.2010.17959.x}, \href
  {http://adsabs.harvard.edu/abs/2011MNRAS.412..979W} {412, 979}

\bibitem[\protect\citeauthoryear{{Weidner}, {Ferreras}, {Vazdekis}  \& {La
  Barbera}}{{Weidner} et~al.}{2013}]{Weidner13}
{Weidner} C.,  {Ferreras} I.,  {Vazdekis} A.,   {La Barbera} F.,  2013, \mn@doi
  [\mnras] {10.1093/mnras/stt1445}, \href
  {http://adsabs.harvard.edu/abs/2013MNRAS.435.2274W} {435, 2274}

\bibitem[\protect\citeauthoryear{{Woosley} \& {Weaver}}{{Woosley} \&
  {Weaver}}{1995}]{WW95}
{Woosley} S.~E.,  {Weaver} T.~A.,  1995, \mn@doi [\apjs] {10.1086/192237},
  \href {http://adsabs.harvard.edu/abs/1995ApJS..101..181W} {101, 181}

\bibitem[\protect\citeauthoryear{{Wuyts} et~al.,}{{Wuyts}
  et~al.}{2010}]{Wuyts10}
{Wuyts} E.,  et~al., 2010, \mn@doi [\apj] {10.1088/0004-637X/724/2/1182}, \href
  {http://adsabs.harvard.edu/abs/2010ApJ...724.1182W} {724, 1182}

\bibitem[\protect\citeauthoryear{{Yan}, {Jerabkova}  \& {Kroupa}}{{Yan}
  et~al.}{2017}]{Yan17}
{Yan} Z.,  {Jerabkova} T.,   {Kroupa} P.,  2017, \mn@doi [\aap]
  {10.1051/0004-6361/201730987}, \href
  {https://ui.adsabs.harvard.edu/abs/2017A&A...607A.126Y} {607, A126}

\bibitem[\protect\citeauthoryear{{Yan}, {Jerabkova}, {Kroupa}  \&
  {Vazdekis}}{{Yan} et~al.}{2019}]{Yan19}
{Yan} Z.,  {Jerabkova} T.,  {Kroupa} P.,   {Vazdekis} A.,  2019, \mn@doi [\aap]
  {10.1051/0004-6361/201936029}, \href
  {https://ui.adsabs.harvard.edu/abs/2019A&A...629A..93Y} {629, A93}

\bibitem[\protect\citeauthoryear{{Yee}, {Ellingson}, {Bechtold}, {Carlberg}  \&
  {Cuillandre}}{{Yee} et~al.}{1996}]{Yee96}
{Yee} H.~K.~C.,  {Ellingson} E.,  {Bechtold} J.,  {Carlberg} R.~G.,
  {Cuillandre} J.~C.,  1996, \mn@doi [\aj] {10.1086/117916}, \href
  {https://ui.adsabs.harvard.edu/abs/1996AJ....111.1783Y} {111, 1783}

\bibitem[\protect\citeauthoryear{{Zhang}, {Romano}, {Ivison}, {Papadopoulos}
  \& {Matteucci}}{{Zhang} et~al.}{2018}]{Zhang18}
{Zhang} Z.-Y.,  {Romano} D.,  {Ivison} R.~J.,  {Papadopoulos} P.~P.,
  {Matteucci} F.,  2018, \mn@doi [\nat] {10.1038/s41586-018-0196-x}, \href
  {https://ui.adsabs.harvard.edu/abs/2018Natur.558..260Z} {558, 260}

\bibitem[\protect\citeauthoryear{{van den Hoek} \& {Groenewegen}}{{van den
  Hoek} \& {Groenewegen}}{1997}]{VDH97}
{van den Hoek} L.~B.,  {Groenewegen} M.~A.~T.,  1997, \mn@doi [\aaps]
  {10.1051/aas:1997162}, \href
  {http://adsabs.harvard.edu/abs/1997A%26AS..123..305V} {123, 305}

\makeatother
\end{thebibliography}
\bibliographystyle{mnras}


\appendix


\bsp	
\label{lastpage}
\newpage

\end{document}